\documentclass[aps,prd,superscriptaddress, twocolumn,10pt]{revtex4-1}
\usepackage{eurosym}
\usepackage{graphicx}
\usepackage{amsmath}
\usepackage{amssymb}
\usepackage{amsfonts}
\usepackage{bm}
\usepackage{float}
\usepackage{color}
\usepackage{multirow}
\usepackage{hhline}

\def\be{\begin{equation}}
\def\ee{\end{equation}}
\def\bea{\begin{eqnarray}}
\def\eea{\end{eqnarray}}

\newcommand{\f}[2]{\frac{#1}{#2}}

\begin{document}

\title{Compact stars in the Einstein dark energy model}
\author{Zahra Haghani}
\email{z.haghani@du.ac.ir}
\affiliation{School of Physics, Damghan University, Damghan, 36716-41167, Iran}
\author{Tiberiu Harko}
\email{tiberiu.harko@aira.astro.ro}
\affiliation{Department of Theoretical Physics, National Institute of Physics
and Nuclear Engineering (IFIN-HH), Bucharest, 077125 Romania}
\affiliation{Astronomical Observatory, 19 Ciresilor Street,
	Cluj-Napoca 400487, Romania,}
\affiliation{Department of Physics, Babes-Bolyai University, Kogalniceanu Street,
	Cluj-Napoca 400084, Romania,}
\affiliation{School of Physics, Sun Yat-Sen University, Guangzhou 510275, People's
	Republic of China,}

\begin{abstract}
We investigate the properties of high density compact objects in a vector type theory, inspired by Einstein's 1919 theory of elementary particles, in which Einstein assumed that elementary particles are held together by gravitational as well as electromagnetic type forces. From a modern perspective, Einstein's theory can be interpreted as a vector type model, with the gravitational action constructed as a linear combination of the Ricci scalar, of the trace of the matter energy-momentum tensor, and of a massive self-interacting vector type field. To obtain the properties of stellar models we consider the field equations for a static, spherically symmetric system, and we investigate numerically their solutions for different equations of state of quark and neutron matter, by assuming that the self-interaction potential of the vector field either vanishes, or is quadratic in the vector field potential. We consider quark stars described by the MIT bag model equation of state, and in the Color Flavor Locked (CFL) phase, as well as compact stars consisting of a Bose-Einstein condensate of neutron matter, with neutrons forming Cooper pairs. Constant density stars, representing a generalization of the interior Schwarzschild solution of general relativity, are also analyzed. As an example of stars described by equations of state obtained by using effective nuclear interactions of the Skyrme type we consider the Douchin-Haensel (SLy) equation of state. The numerical solutions are explicitly obtained in both standard general relativity, and the Einstein dark energy model, and an in depth comparison between the astrophysical predictions of these two theories is performed. As a general conclusion of our study we find that for all the considered equations of state a much larger variety of stellar structures can be obtained in the Einstein dark energy model, including classes of stars that are more massive than their general relativistic counterparts. As a concrete application of our results we suggest that compact objects with masses of the order of $2.5\,M_{\odot}$, associated, for example, with the GW 190814 gravitational wave event, could be in fact quark or neutron stars, described by the Einstein dark energy model.
\end{abstract}

\maketitle

\tableofcontents

\section{Introduction}

The maximum mass of a neutron star is a question of fundamental importance from both theoretical and observational points of view. An upper limit for the critical mass of a star $M_{max}$, of the order of $3.2\,M_{\odot}$, was found in \cite{Ru} from the analysis of the general relativistic hydrostatic equilibrium Tolman-Oppenheimer-Volkoff equation, by considering the limiting case of the stiff fluid equation of state of the dense matter. Still, in this bound the effects of the rotation and of the existence of exotic states of matter are ignored.  On the other hand, a number of theoretical arguments, that seemed to be supported by the observational evidence, suggested that
neutron stars had a characteristic, unique mass of the order of $1.4\,M_{\odot}$ \cite{Ho}. Assuming an interaction among neutrons, the role of the repulsive interactions should be important, thus leading to a numerical coincidence with the Chandrasekhar mass \cite{Sha}. This coincidence led to the postulation that some specific physical processes at the birth of neutron star would uniquely fix its mass \cite{Ho}. However, with the increase in the precision of the astronomical observations, the paradigm of a unique mass distribution around $1.4\,M_{\odot}$ of the neutron stars has to abandoned, and the existence of neutron stars with light $\left(1.174 \pm 0.004\,M_{\odot}\right)$ \cite{Ma}, or heavy $\left(2.140.010.09\,M_{\odot}\right)$ \cite{Cro} masses is presently well established. These masses are very different from the long time assumed Chandrasekhar mass limit. Moreover, in \cite{Valentim:2011vs}, it was shown, by using 54 measured values, that the neutron star mass distribution can be represented by bimodal distribution, with the first peak located at 1.37 $M_{\odot}$, with a much larger second peak appearing at 1.73 $M_{\odot}$.

A new window on the gravitational processes that play a fundamental role in many astrophysical processes was opened by the experimental detection of gravitational waves by the LIGO and VIRGO scientific collaborations \cite{Abbott:2016blz,TheLIGOScientific:2016wfe}. The experimental study of the gravitational waves leads to a better understanding of the properties of compact objects, including the mass distribution of the neutron stars. Another very important advance in gravitational physics and astrophysics is represented by the GW170817 event \cite{TheLIGOScientific:2017qsa}. This experimental event initiated the multimessenger Era, with the signal detected worldwide by more than 60 instruments. In GW170817 the gravitational wave is produced by the merging of two neutron stars, and it originates from the shell elliptical galaxy NGC 4993. The GW170817 event implies a mass of the nonrotating neutron star of the order of $M\leq 2.3\,M_{\odot}$ (see \cite{Baiotti:2016qnr} for a review of the merger of binary neutron stars). The merger of neutron stars takes place in conditions of very high gravitational fields, and it leads to the emission of intense
fluxes of gravitational waves that can be detected experimentally.

The experimental study of the gravitational waves also led to some intriguing results that are likely to modify a number of
basic paradigms in present day astrophysics. The
GW190814 event \cite{Abbott:2020khf} indicates an unusual structure of the
mass components of the merging stars, with one of the masses having values of the order of $2.5 - 2.6\,M_{\odot}$
(90\% confidence). For this gravitational event no optical counterpart was detected. If the observed object is a neutron star, its high mass
would contradict the paradigm of the existence of a $%
1.4\,M_{\odot}$ mass scale for neutron stars, or of the mass distributions found in \cite{Valentim:2011vs}, with the mass observed in the GW190814 event
located far away from the previously inferred peaks.

An accurate measurement of the mass of the millisecond
pulsar MSP J0740+6620, using Shapiro delay, gave another intriguing value, namely $2.14 ^{+ 0.10}_{ -0.09}\,M_{\odot}$ \cite{Cromartie:2019kug}. Several other similarly high mass values have also been measured. Recently,  a companion of V723 Mon,  a nearby red giant, having a mass of around $3\,M_{\odot}$, has been observed \cite{BHm}.  This discovery also leads to new questions on the mass distribution of neutron stars, and on the formation of black holes from the collapse of massive objects. In fact, it was found that the masses of neutron stars in gravitational-wave binaries are consistent with a uniform distribution, with a greater
prevalence of high-mass neutron stars \cite{Lan}.

These observations  of the masses of neutron stars require some drastic modifications in our understanding of the structure of compact objects, since in order to
explain the observed values we need either to assume that the density is much higher than the nuclear saturation density, leading to a significant modification of the equation of state of dense matter, or to assume that at high densities the gravitational force itself changes its behavior.

There are a number of physical or astrophysical effects that could lead to the increase of the masses
of neutron stars, and thus explain the GW190814 event. The information obtained from the GW170817 event was used in \cite{Fer} to make a probabilistic inference of the equation of state of dense stars, which goes beyond the constraints imposed by nuclear matter properties, which do not allow one to distinguish between equations of state that predict different neutron star maximum masses. Constraints on the dense matter equation of state and neutron star properties from PSR J0740+6620 and multimessenger observations were obtained in \cite{Raa}. An analysis of an updated sample of neutron star masses, derived from the study of a variety of 96 binary systems containing at least one neutron star, and performed in \cite{Ho1} led to the conclusion that the maximum mass implied by the sample is of the order of $2.5-2.6\,M_{\odot}$. A Bayesian analysis of the maximum mass of neutron stars with a quark core, incorporating the observational data from tidal deformability of the GW170817 binary neutron star merger as detected by LIGO/Virgo and the mass and radius of PSR J0030+0451  gave an absolute upper bound around $2.85\,M_{\odot}$ for the mass of the star \cite{Li}. The equation of state of dense matter, up to twice nuclear saturation density, obtained from chiral effective field theory, and the recent observations of neutron stars were used in \cite{Dri} to gain some insights about the high-density matter. A joint Bayesian inference of neutron-star mass and radius constraints based on GW170817, observations of quiescent low-mass X-ray binaries (QLMXBs), photospheric radius expansion X-ray bursts (PREs), and X-ray timing observations of J0030+0451, was performed in \cite{AlM},  indicating that the gravitational-wave and electromagnetic observations of neutron-star structure can provide a consistent picture of the neutron-star mass-radius curve, and of the equation of state. The question of the maximum mass of neutron stars was reconsidered in \cite{Go}, by using a Markov chain Monte Carlo approach to generate about 2 million phenomenological equations of state, with and without first order phase transitions. The impact of a 2.6 solar mass neutron star on the nucleonic equations of state was considered in \cite{Fat}. In \cite{Mos} it was shown that in the recent detection of GW190814, representing the merger of a binary with a primary having a mass of $\sim 23\,M_{\odot}$, and a secondary with a mass of $\sim 2.6\,M_{\odot}$, the secondary could be interpreted as either the lightest black hole, or the most massive neutron star ever observed. However, it could also be an indication of a novel class of exotic compact objects. It was also proposed that GW190814 is the result of
the merging of a black hole -- strange quark star system \cite{Bombaci:2020vgw,Wu:2020zhr,Horvath:2020lwj}.

If a phase transition from neutron to quark matter takes place in the dense matter inside a neutron star, the physical properties of the newly formed object may be very different as compared to the properties of standard neutron stars. Stellar mass black holes, with masses in the
range of $3.8\,M_{\odot}$ and $6\,M_{\odot}$, may be quark stars in
the Color-Flavor-Locked (CFL) phase, as suggested in \cite{Kovacs:2009kv}. Rotating CFL
quark stars may have much higher masses than ordinary neutron stars.  One can distinguish quark stars in the CFL or standard phases from low mass black
holes or neutron stars through the comparative study of thin accretion disks that form around these types of objects, and Kerr black holes,
respectively \cite{Ko1}. A significant part of the matter component inside a neutron star may exist in the form of a Bose-Einstein condensate (BEC) \cite{Chavanis:2011cz}. The  astrophysical parameters of the neutron stars containing matter in the form of a BEC strongly depend on the
mass of the condensate particle, and on its scattering length. One can conjecture that neutron stars with masses in the range of $2-2.5\, M_{\odot} $
could be in fact BEC stars, containing a significant amount of matter in a condensate phase \cite{Chavanis:2011cz}. The properties of condensate stars can also be investigated via the electromagnetic emissions from their accretion disks \cite{Da1}.

Another promising avenue for the explanation of the high masses of some neutron
stars is represented by the possible modification of the nature and characteristics of the
gravitational interaction at very high densities. This would require the description of the structure of
neutron stars in the framework of modified theories of gravity. A classic result in general relativity is the Buchdahl limit \cite{Bu}, which states that for stable compact objects the mass-radius ratio must satisfy the constraint $2M/R\leq 8/9$. However, various geometrical and physical effects can modify the Buchdahl bound. For example,
in the presence of a cosmological constant $\Lambda$, for the mass-radius $M/R$ ratio neutron stars we obtain a constraint of the form $2M/R\leq \left(1-8\pi \Lambda R^2/3
\right)\left[1-\left(1-2\Lambda/\bar{\rho}\right)^2/9\left(1-8\pi \Lambda
R^2/3\right)\right]$ , where $\bar{\rho}$ is the mean density of the star \cite{Mak:2001gg}. In modified gravity theories
in which an effective contribution $\theta
_{\mu}^{\nu}$ to the matter energy-momentum tensor $T_{\mu}^{\nu}$ does appear, the Buchdahl limit is also modified.  By defining an effective
density $\rho _{\mathrm{eff}}c^{2} = \rho c^{2}/G
+ \theta _{0}^{0}$, and an effective mass $m_{\mathrm{eff}}=4\pi \int _0^r{%
r^2\rho _{\mathrm{eff}}dr}$, the generalized Buchdahl bound
in modified gravity theories can be obtained as \cite{Burikham:2016cwz}
\begin{equation}  \label{58}
\frac{2m_{\mathrm{eff}}(r)}{r}\leq 1-\left[1+\frac{2\left(1+f(r)\right)}{%
1+4\pi w_{\mathrm{eff}}(r)}\right]^{-2},
\end{equation}
where
\begin{equation}  \label{f(r)*}
f(r) = 4\pi \frac{\Delta (r)}{\langle \rho_{\mathrm{eff}}\rangle (r)}\left\{%
\frac{\arcsin\left [\sqrt{2m_{\mathrm{eff}}(r)/r}\right ]}{\sqrt{%
2m_{\mathrm{eff}}(r)/r}} -1\right \},
\end{equation}
$w_{\mathrm{eff}}(r)=p_{\mathrm{eff}}/\langle \rho_{\mathrm{eff}}\rangle (r)$%
, and $\Delta =\left(G/c^4\right)\left(\theta _1^1-\theta _2^2\right)$,
respectively. Therefore, these general considerations indicate that the supplementary contributions to the matter energy-momentum
tensor resulting from the modifications of the gravitational field equations generally lead to the increase of the mass of the compact object.
The physical properties and structure of the neutron, quark and other types of exotic stars were in different modified gravity theories in \cite{Star1,Star2,Star3,Star4,Star5,Star6,Star7,Star8,Star9,Star10,Star11}.  The upper mass limit predictions of the baryonic mass for static neutron stars in the context of  $R^2$ gravity were investigated in \cite{Star10}. The maximum baryonic mass of static neutron stars was calculated by adopting several realistic equations of state as well as the stiff matter equation of state. It was found that maximum neutron star masses are likely to be in the lower limits of the range of $M\sim 2.4-3M_{\odot}$, and that neutron stars cannot have gravitational masses larger than $3M_{\odot}$.

An interesting, but forgotten and underrated modified gravity theory was proposed by Einstein soon after the birth of general relativity.
After constructing a static model of the Universe, which required the addition of
the cosmological constant in the gravitational field equations \cite{Ein1},
Einstein addressed the problem of the structure of the elementary
particles \cite{Ein2}. By assuming that the fundamental forces acting on elementary particles
are the gravitational force, described by the metric
tensor $g_{\mu \nu}$ and its derivatives, and the electromagnetic forces,
with energy-momentum tensor  $S_{\mu \nu}$, obtained from
the electromagnetic fields $F_{\mu \nu}$,  Einstein proposed the basic equation describing the microscopic world as \cite{Ein2}
\begin{equation}  \label{1}
R_{\mu \nu}+\bar{\lambda}g_{\mu \nu }R=\kappa ^2 S_{\mu \nu},
\end{equation}
where $\kappa ^2 =8\pi G/c^4$ is the
gravitational coupling constant, and $\bar{\lambda}$ is a constant. Taking into account that  $S_{\mu}^{\mu}=0$, one can determine  $\bar{\lambda}$
from the trace of Eq.~(\ref{1}) as $\bar{\lambda}%
=-1/4 $. Thus Einstein's equation (\ref{1}) can be written as
\begin{equation}  \label{2}
R_{\mu \nu}-\frac{1}{4}\,g_{\mu \nu }R=\kappa ^2 S_{\mu \nu}.
\end{equation}

By assuming that the gravitational field equations in the presence of a cosmological equation are still valid, Einstein determined the matter energy-momentum tensor,  and reformulated the field equation (\ref{2}) as
\begin{equation}  \label{10}
R_{\mu \nu}-\frac{1}{4}g_{\mu \nu}R=\kappa ^2 \left(T_{\mu \nu}-\frac{1}{4}%
Tg_{\mu \nu}\right).
\end{equation}

The field equations (\ref{10}) can be called the \textit{%
geometry-matter symmetric Einstein equations}. Einstein's approach to provide a geometric solution to the problem of the structure of the matter attracted very little
interest  \cite{p1,p2}.  It was however briefly mentioned  as a
possible solution to the cosmological constant problem, since it follows from geometry-matter symmetric Einstein equations as an
integration constant \cite{Wein}.

Rastall \cite{Rastall} proposed a theory somehow similar to Einstein's theory of elementary particles, in which the electromagnetic
energy-momentum tensor in Eq.~(\ref{1}) is replaced by the ordinary
matter energy-momentum tensor, $S_{\mu \nu}\rightarrow T_{\mu \nu}$, with the field equations given by
\begin{equation}
R_{\mu \nu}+\bar{\lambda}g_{\mu \nu }R=\kappa ^2 T_{\mu \nu}.
\end{equation}
 Hence, the matter
energy-momentum tensor is not conserved, and $\nabla
_{\mu}T_{\nu}^{\mu}=\lambda \nabla _{\nu}R$, $\lambda =\mathrm{constant}$.
However,  in \cite{Ri} it was shown that
the Rastall theory is just a particular case of the $f(R,T)$ gravity theory
\cite{frt1}, where $T$ is the trace of the energy-momentum tensor, a modified gravity theory that is based on the existence of non-minimal curvature-matter coupling.

Einstein's theory of elementary particles was reinterpreted as a vector type dark energy model in \cite{EDE},  by considering  a gravitational action containing a linear combination of the Ricci scalar $R$,  and of the trace of the ordinary matter energy-momentum tensor $T$.  The existence of a massive self-interacting vector type dark energy field, coupled with the matter current was also assumed. In this model the matter energy-momentum tensor is not conserved, and thus the resulting gravitational field equations can also be interpreted by using the formalism of the thermodynamics of open systems as describing particle generation from the gravitational field. In the vacuum case the model admits a de Sitter type solution. The cosmological parameters, including Hubble function, deceleration parameter and matter energy density have been obtained as a function of the redshift by using analytical and numerical techniques, and for different values of the model parameters. In \cite{EDE} it was shown that for all considered cases the Universe experiences an accelerating cosmological expansion, ending with a de Sitter type evolution. Also, in \cite{EDE-pert} the growth rate of matter perturbations in the Einstein dark energy theory is considered. The dynamical system analysis of this model has shown that there are three fixed points corresponding to the dust, radiation and de Sitter universes. In \cite{EDE-pert}, the model parameters are fitted with observational data, using two independent datasets corresponding to the Hubble parameter $H$ and also $\sigma_8$. The theory is then shown to be consistent with observational data.

It is the goal of the present paper to investigate the properties of relativistic compact high density objects in the generalized vector-tensor version of the Einstein dark energy model \cite{EDE}. To study the interior solutions of the field equations we adopt a static spherically symmetric geometry for the star, and we assume that the matter content is represented by a perfect fluid. Then, after writing down the gravitational field equations, as a first step in our study we derive the mass continuity equation and the Tolman-Oppenheimer-Volkoff equation, which describe, together with the equation of motion of the vector field,  the basic astrophysical parameters (mass and radius)
of the star. The stellar structure equations in the Einstein dark energy model are then solved numerically for
several equations of state of the dense matter. As for the self-interaction potential of the vector field we assume that it either vanishes, or is quadratic in the vector field potential.

As a first specific example of a high density compact object in the theory, we consider the case of the constant density stars. Even though this kind of object is not considered appropriate for a realistic description of stellar objects, they still have a major theoretical importance.  As a second class of stars we consider quark stars in both standard phase, described by the MIT bag model equation of state, and in the Color-Flavor-Locked phase, in which the quarks form Cooper pairs, whose color properties are correlated with their flavor properties in a one-to-one correspondence between three color pairs and three flavor pairs. The Color-Flavor-Locked phase is the highest-density pase of three-flavor colored matter. The
Bose-Einstein condensate equation of state, corresponding to a polytropic equation of state with polytropic index
$n = 1$ is considered within the framework of the Einstein dark energy model. Finally, we will consider stars described by using effective nuclear interactions of the Skyrme type (SLy). For all these matter equations of state the global astrophysical parameters of the compact objects (radius and mass), as well as the vector field, are obtained in both standard general relativity and the Einstein dark energy model. This approach
allows us to perform an in depth comparison of the two theories that could be used for the description of stellar properties,  and internal structure. As a
general conclusion of our investigations we find that the Einstein dark energy model permits the existence of more massive stable stellar objects,
as compared to standard general relativity.

The present paper is organized as follows. The variational principle of the
Einstein dark energy model is introduced in Section~\ref{sect2}, where the
corresponding field equations are also presented. The spherically symmetric static gravitational field equations are also written down, and the generalized Tolman-Oppenheimer-Volkoff and mass continuity equations are derived. In Section~\ref{sect3} several stellar models, corresponding to different equations of state of the dense matter, are investigated, including constant density stars, quark stars in normal and CFL phase, as well as Bose-Einstein condensate and Sly stars, for two particular forms of the vector field self-interaction potential. We discuss and conclude our results in
Section~\ref{sect4}.

\section{Compact stars in the Einstein dark energy model}\label{sect2}

The action and field equations of Einstein dark energy model have been introduced in \cite{EDE}. In this approach to the dark energy problem we assume that the Universe is filled with a cosmological dark energy vector field $\Lambda _{\mu }\left( x^{\nu
}\right) $. We define the dark energy strength tensor according to
\begin{equation}
C_{\mu \nu }=\nabla _{\mu }\Lambda _{\nu }-\nabla _{\nu }\Lambda _{\mu }.
\end{equation}%

In the following we investigate the local effects of such a vector field, which may not be necessarily related to the cosmological dynamics. We define the energy-momentum tensor $T_{\mu \nu}$ of the baryonic matter fields as
\be
T_{\mu \nu}=-\frac{2}{\sqrt{-g}} \frac{\partial \left( \sqrt{-g} \mathcal{L}_{m} \right)}{%
\partial g^{\mu\nu}},
\ee
where  $\mathcal{L}_{m}$ is the Lagrangian of the ordinary matter.

The Einstein dark energy model  action is given by \cite{EDE}
\begin{align}
S =\int d\,^4x\sqrt{-g}\,\Bigg[\f{1}{2\kappa^2}\left( 1-\beta_1 \right)& R+%
\frac{\beta_2 }{2}\,T-\frac{1}{4}C_{\mu \nu }C^{\mu \nu }  \notag  \label{s1a}
\\
& +V(\Lambda^2)+\mathcal{L}_{m} \Bigg] ,
\end{align}%
where $\beta_1$, $\beta_2$ are two arbitrary dimensionless constants. Also, the potential term $V$ is an arbitrary function of $\Lambda^2=\Lambda_\mu\Lambda^\mu$.

The energy-momentum tensor $S_{\mu \nu }$ of the dark energy field is given by
\begin{equation}  \label{23}
S_{\mu \nu }=C_{\mu \alpha }C_{\nu }^{~\alpha }-\frac{1}{4}g_{\mu \nu
}C_{\alpha \beta }C^{\alpha \beta },
\end{equation}
and it has the property $S_{\mu}^{\mu}=0$.

By varying the action (\ref{s1a}) with respect to the metric gives the field equations
\begin{align}\label{eq1}
&\kappa^2(1-\beta_1)G_{\mu\nu}-\f12S_{\mu \nu}
-\frac{1}{2}g_{\mu\nu}V+\Lambda_\mu\Lambda_\nu V^\prime \nonumber\\
&=\f12(1+\beta_2)T_{\mu\nu}-\f12\beta_2\,\left(\mathcal{L}_m-\frac{1}{2}T\right)g_{\mu\nu},
\end{align}
where a prime denotes derivative with respect to the argument of the function.

By varying the action \eqref{s1a} with respect to the vector potential of the field, we obtain the equation
\begin{equation}\label{lam}
\nabla _{\nu }C^{\mu\nu }=2\Lambda^\mu V^\prime.
\end{equation}

By taking the divergence of the metric field equation \eqref{eq1}, and using equation \eqref{lam}, one obtains the balance equation of the matter energy-momentum tensor as
\begin{align}\label{cons1}
\nabla^\mu T_{\mu\nu}&=\f{\beta_2}{1+\beta_2}\nabla_\nu\left(\mathcal{L}_m-\f12T\right).
\end{align}

\subsection{Field equations for high density static spherically symmetric objects}

In order to investigate the properties of dense stars in the Einstein dark energy model, we adopt for the line element the standard static, spherically symmetric form, given by
\begin{align}
ds^2=-e^{-2f(r)}dt^2+\f{1}{1-2m(r)/r}dr^2 +r^2 d\Omega^2,
\end{align}
where the two metric tensor components $g_{tt}=-e^{-2f(r)}$ and $g_{rr}=1/(1-2m(r)/r)$ are functions of the radial coordinate $r$ only.
As for the vector field, we adopt the ansatz
\begin{align}
\Lambda^\mu=\sqrt{|g^{tt}|}h(r)\delta^\mu_t,
\end{align}
where $h(r)$ is an arbitrary function to be determined from the field equations.  The energy-momentum tensor of the matter is given by
\begin{align}
T_{\mu\nu}=\left(p+\rho \right)u_\mu u_\nu +p\, g_{\mu\nu},
\end{align}
where $\rho$ and $p$ are the energy density, and pressure, respectively.

With these assumptions, the vector field equation \eqref{lam} becomes
\begin{align}
&h'' \left(1-\frac{2 m}{r}\right)+h' \left[\f{m}{r^2} \left(2 r f'-3\right)- f'-\f{m'-2}{r}\right] \nonumber\\&+h
	\left[  \f{m'-2}{r}f'+\f{3 m}{r^2} f'- f'' \left(1-\f{2 m}{r}\right)-2
	V'\right]=0,
\end{align}
where  $V(\Lambda^2)=V(-h^2)$.

The balance equation of the energy-momentum tensor Eq.~\eqref{cons1} gives
\begin{align}\label{pprime}
2p'-\beta_2\left( \rho'+p'\right)=2\left(1+\beta_2\right)\left(\rho+p\right)f'.
\end{align}

The components of the metric field equations are then as follows (in the following we set $\kappa^2 =1$)
\begin{align}
4&\left(1-\beta _1\right)\frac{
	m'}{ r^2}=2\rho+\beta _2\left(\rho-3p\right)+\rho_\Lambda,
\end{align}
\begin{align}\label{f2}
4\left(1-\beta _1\right)\left[\f{m}{r^3}+\f{f'}{r}\left(1-\f{2m}{r}\right)\right]=-2p-\beta_2(5p+\rho)+p_\Lambda,
\end{align}
\begin{align}
2&\left(1-\beta _1\right)\bigg[\left(1-\f{2m}{r}\right)\left(f''-f'^{\,2}\right)+\left(m'-\f{m}{r}\right)\f{f'}{r}\nonumber\\&
+\f{1}{r^2}\left(m'-\f{m}{r}\right)\bigg]=-2p-\beta_2(5p+\rho)+p_\Lambda,
\end{align}
where we have denoted
\be
\rho_\Lambda =\left(1-\f{2 m}{r}\right)\left(h'^{\,2}-2\,h
\, h'\,  f'+h^2 f'^{\,2}\right)+ 2 V	+4 h^2 V',
\ee
and
\be
p_\Lambda=\rho_\Lambda-4h^2 V',
\ee
respectively.

In the following we will assume that the dense matter satisfies a barotropic equation of state, $p=p(\rho)$. Hence, $dp/dr=(dp/d\rho)(d\rho /dr)$, giving $d\rho /dr=\left(1/c_s^2\right)(dp/dr)$, where $c_s^2=dp/d\rho$ is the speed of the sound. Hence, Eq.~(\ref{pprime}) can be written as
\be\label{p1}
p'=-\frac{\left(1+\beta _2\right)\left(\rho+p\right)f'}{\left(\beta _2/2\right)\left(1+1/c_s^2\right)-1}.
\ee

Since the pressure must be a monotonically decreasing function inside the star, the condition $p'<0$ must hold $\forall r\in[0,R)$, and thus the coupling parameter $\beta _2>0$ must satisfy the constraint
\be
\frac{\beta _2}{2}\left(1+\frac{1}{c_s^2}\right)>1.
\ee

By expressing $f'$ from Eq.~(\ref{f2}), and after substituting it in Eq.~(\ref{p1}), we obtain the generalized Tolman-Oppenheimer-Volkoff (TOV) equation in the Einstein dark energy model as
\begin{widetext}
\be
\frac{dp(r)}{dr}=-\frac{\left( 1+\beta _{2}\right) \left( \rho (r)+p(r)\right)
\left\{ \left[ p_{\Lambda }-2p(r)-\beta _{2}\left( 5p(r)+\rho (r) \right) \right]
r^{3}-4\left( 1-\beta _{1}\right) m(r)\right\} }{4\left( 1-\beta _{1}\right) %
\left[ \left( \beta _{2}/2\right) \left( 1+1/c_{s}2\right) -1\right] r^{2}\left(1-2m(r)/r\right)}.
\ee
\end{widetext}

The TOV equation must be solved together with the mass continuity equation, which takes the form
\be
\frac{dm(r)}{dr}=\frac{1}{4\left(1-\beta _1\right)}\left[2\rho (r)+\beta _2\left(\rho (r)-3p(r)\right)+\rho_\Lambda\right]r^2,
\ee
and with the boundary conditions $m(0)=0$, $\rho(0)=\rho _c$, and $p(R)=0$, respectively.

In the next sections, we will consider different kinds of equation of states for the compact stars. In order to simplify the mathematical formalism for the study of dense stellar type objects we will use a set of dimensionless parameters, defined as follows,
\begin{align}
& p=\rho_c \,\bar{p}, \quad \rho=\rho_c\, \bar{\rho}, \quad \bar{m}=\sqrt{\rho_c}\, m\nonumber\\& \eta=\sqrt{\rho_c}\,r,  \quad h=\rho_c\, \bar{h}
\end{align}
where $\rho_c$ is the central density of the star.

\subsection{Constraints on the model parameters}

The nature and structure of the stellar models in the Einstein dark energy model essentially depend on the model parameters $\beta _1$ and $\beta_2$, respectively.  There are a number of high precision observational tests that allow to obtain very strong constraints on the free parameters of the gravitational theories. Such tests are represented by the Shapiro delay observations in the case of massive objects \cite{Nat1,Nat2}, which can yield precise mass determinations for both a millisecond pulsar and of its companion star. However, it is important to point out that the delay effect can only be easily observed in a small subset of high-precision, highly inclined (nearly edge-on) binary pulsar systems \cite{Nat2}. By using the Shapiro delay effect the mass of PSR J1614-2230, which show a strong Shapiro delay signature, has been determined to be $1.97\pm 0.04 M_{\odot}$ \cite{Nat1}, while for the mass of the millisecond pulsar MSP J0740+6620 the value  $2 .14_{-0.09}^{ +0 .10} M_{\odot}$ \cite{Nat2} has been obtained. Further important tests of gravity theories can be obtained from the precession of the planet Mercury \cite{Per}, or from the study of the recently detected gravitational waves \cite{GW}.

The allowable range of the parameters $\beta _1$, $\beta _2$ of the Einstein dark energy model, and of the potential of the form $V\left(\Lambda_0\right)=V_0+\nu \Lambda_0^2$, where $V_0$ and $\nu$ are constants, was investigated in the cosmological setting in \cite{EDE}. Since the time variation of the gravitational constant $G$, which also contains the model parameter $\beta _1$ is very small, in the cosmological approach it has been neglected from the field equations, by taking $\beta _1\approx 0$. Secondly, the constant term in the potential can be interpreted as the cosmological constant. As for $\beta _2$, in order to obtain consistency with the cosmological observations at low redshift, it must take values in the range $\beta _2\leq 4$. This high value of $\beta _2$ is not surprising, since in the present model dark energy, representing around 75\% of the energy of the Universe, is generated from ordinary matter (around 25\%, via the coupling $\beta _2T$ in the matter Lagrangian. Hence, in the cosmological approach, the contribution of the term $\beta _2T$, together with that of the vector field potential,  must exceed several times the contribution coming from the matter Lagrangian alone, so that this term can act as dark energy. As for the parameters of the potential $V$, $V_0$ can be interpreted as a cosmological constant, while $\nu$ is determined so that the Universe enters in a de Sitter accelerating phase.

However, the situation is different in the case of high density compact objects. The contribution to the total energy balance of the term $\beta _2T$, containing the trace of the energy-momentum tensor, cannot exceed significantly the contribution of the matter Lagrangian $L_m$, and it gives mostly a gravitational type correction to the energy content of the star. Hence, in the following we will assume that $\beta _2$ is smaller than 1, $\beta_2<1$. Moreover, we will not impose any a priori restrictions on the sign of $\beta_2$, allowing both positive and negative values for it. In vacuum, the gravitational field equations of the Einstein dark energy model reduce to the standard form of a vector-tensor theory,
\be\label{RN}
\kappa^2(1-\beta_1)G_{\mu\nu}-\f12S_{\mu \nu}
-\frac{1}{2}g_{\mu\nu}V+\Lambda_\mu\Lambda_\nu V^\prime =0.
\ee

If the vector field potential can be neglected, the vacuum spherically symmetric solution of Eq.~(\ref{RN}) is given by the effective Reissner-Nordstr\"{o}m metric, with $g_{tt}=-\left(1-2M/r+Q/r^2\right)$, and $g_{rr}=1/\left(1-2M/r+Q/r^2\right)$, respectively, where $Q=Q_0/\kappa ^2\left(1-\beta _1\right)$, with $Q_0$ an integration constant, plays the role of an effective charge. The effective charge $Q$ can be constrained from the perihelion precession of Mercury as \cite{BoHa10}
\be
\left|Q\right|\leq \frac{M_{\odot}a\left(1-e^2\right)}{\pi} \Delta \phi,
\ee
where $a$ is the semi-major axis of the planet, $e$ the eccentricity of the orbit, and $\Delta \phi=\delta \phi_{Obs}-\delta \phi_{GR}=0.17\pm 0.21$ arcsec per century is the excess perihelion precession that cannot be explained by general relativity. By taking into account the observational data for Mercury \cite{Per}, we obtain $\left|Q\right|\leq \left(5.17\pm 6.39\right)\times 10^4$ m$^2$, or $\left|Q\right|\leq \left(1.32\pm 1.63\right)\times 10^{30}$ MeV$^{-2}$. Hence for $\beta _1$ we obtain the expression
\be
\beta _1\leq 1-\frac{\pi Q_0}{\kappa ^2M_{\odot}a\left(1-e^2\right)\Delta \phi}.
\ee

Hence the value of $\beta _1$ depends on the dark energy charge $Q_0$ of the central object. If the second term in the above equation can be neglected, then the restriction for $\beta _1$ reduces to $\beta_1\leq 1$, a condition required to assure the positivity of the gravitational coupling. However, in the following we will assume that the dark energy charge $Q_0$ of the dense compact objects is high, and we will assume that $\beta _1$ can take positive values in the range $\beta _1\leq 0.10$, and we will also allow negative values for it.

\section{Dense stellar type structures in the Einstein dark energy model}\label{sect3}

In the following we will consider the structure and astrophysical properties of several classes of compact objects in the Einstein dark energy model. In particular, we will investigate constant density stars, two types of quark stars, with the first, described by the MIT bag model equation of state, consisting of a mixture of $u$, $d$ and $s$ quarks.  At ultra-high
densities, quark matter may exist in a variety of superconducting states, in the so-called Color-Flavor-Locked (CFL) phase. An interesting class of objects is represented by the Bose-Einstein condensate stars, in which it is assumed that the dense matter underwent a phase transition to form a Bose-Einstein cFigsondensate. For each case we will consider two forms of the vector field potential, corresponding to the cases $V=0$, and to the Higgs type $V=\lambda +a\Lambda _{\mu}\Lambda ^{\mu}$ \cite{Aad}, respectively, where $a$ and $\lambda$ are constants.

\subsection{Constant density stars}

The first interior solution of the Einstein gravitational field equations was obtained by Schwarzschild in 1916 \cite{Sch} under the assumption of the constant density of the star. It has a remarkable mathematical simplicity, and despite the fact that it is generally considered as not providing a realistic description of neutron stars, its properties have been intensively investigated \cite{Mah}. In particular, for constant density stars the Buchdahl bound becomes exact, so that $2M/R=8/9$.

We begin our study of the compact objects in the Einstein dark energy model by considering the case in which the density is constant throughout the star. In the following we study the numerical solution of the model for $\rho={\rm constant}=2.15\times 10^{14}\;{\rm g/cm}^3$.

\subsubsection{The case $V=0$}

\begin{figure*}[htbp]
	\centering
	\includegraphics[width=8.0cm]{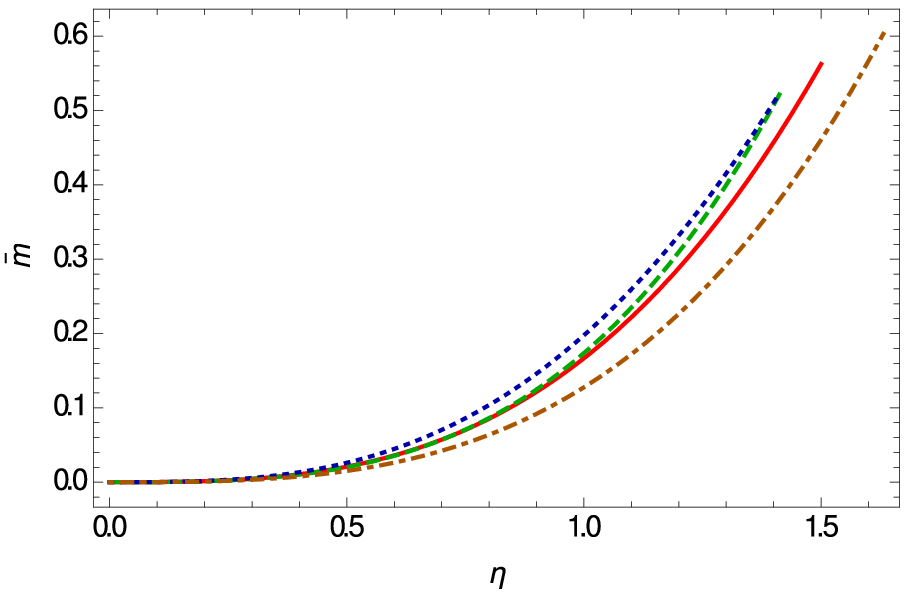}\hspace{.4cm}
	\includegraphics[width=8.0cm]{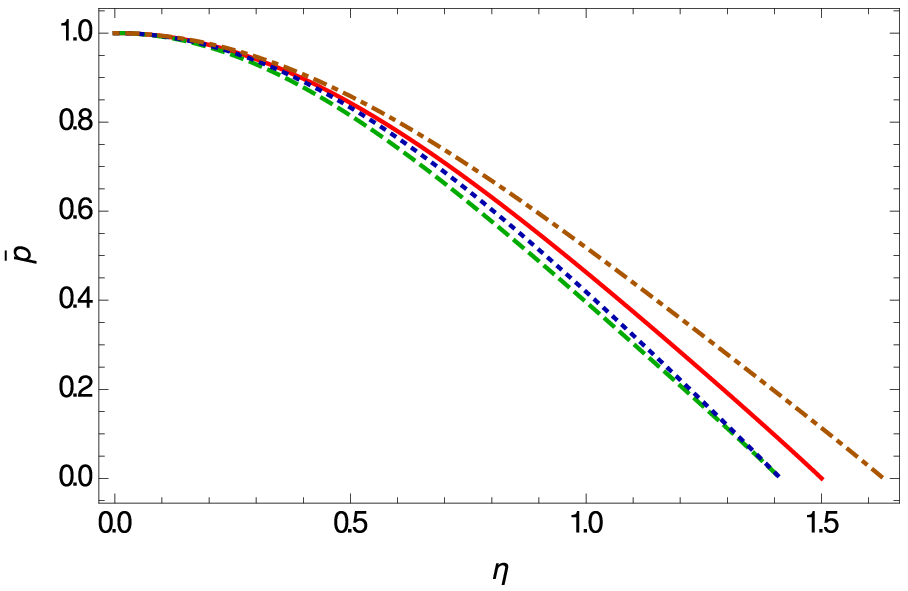}
	\caption{Variation of the interior mass and pressure profiles of a constant density star as a function of the dimensionless radial distance $\eta$ from the center of the  star for $V=0$, and for three different values of the
		constants $\beta_1$ and $\beta_2$: $\beta_1=0.10$  and  $\beta_2=0.15$  (dashed curve),  $\beta_1=0.10$  and  $\beta_2=-0.15$  (dotted curve), and  $\beta_1=-0.20$  and  $\beta_2=0.15$ (dot-dashed curve).  For the central pressure of the star we have adopted the value $p_{c}=1.93\times 10^{35} {\rm erg/cm}^3$, while $\bar{h}_0=0.1$ and $\bar{h}^\prime_0=0.5$. The solid curve represents the standard general relativistic mass and pressure profile. }
	\label{consdens-mass-dens-v0}
\end{figure*}

\begin{figure}[htbp]
	\centering
	\includegraphics[width=8.0cm]{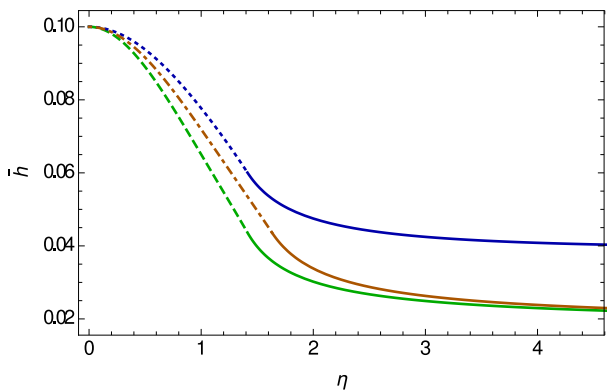}
	\caption{Variation of the temporal component of dark energy vector field inside and outside (solid lines) of the star with constant density as a function of the radial distance from the center of the  star $\eta$ for $V=0$, and for three different values of the
		constants $\beta_1$ and $\beta_2$: $\beta_1=0.10$  and  $\beta_2=0.15$  (dashed curve),  $\beta_1=0.10$  and  $\beta_2=-0.15$  (dotted curve), and  $\beta_1=-0.20$  and  $\beta_2=0.15$ (dot-dashed curve).  For the central pressure of the star we have adopted the value $p_{c}=1.93\times 10^{35} {\rm erg/cm}^3$, while $\bar{h}_0=0.1$ and $\bar{h}^\prime_0=0.5$.  }
	\label{consdens-vec-v0}
\end{figure}

The behaviors of mass, pressure and of the non-zero component of the vector dark energy $\bar{h}$ inside the star are presented in Figs.~\ref{consdens-mass-dens-v0} and \ref{consdens-vec-v0}, respectively.

As one can see from Figs.~\ref{consdens-mass-dens-v0}, the mass profile is an increasing function of $\eta$, while the pressure is a monotonically decreasing function, vanishing on the surface of the star. In Fig.~\ref{consdens-vec-v0} is shown that the dimensionless vector field is a monotonically decreasing function of $\eta$ inside the star and outside the star also decreases and asymptotically reaches to a constant value. All physical parameters are strongly dependent on the model parameters $\beta _1$ and $\beta _2$.

\begin{figure}[htbp]
	\centering
	\includegraphics[width=8.0cm]{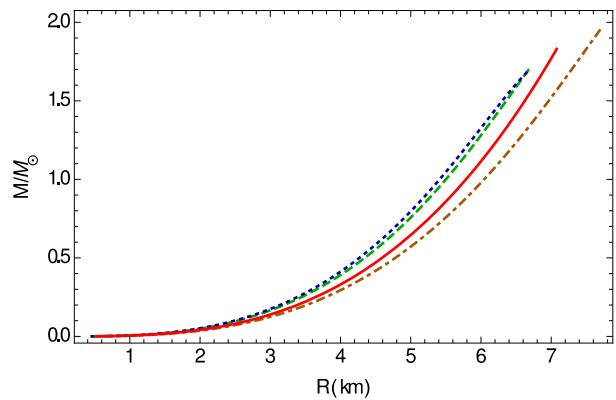}
	\caption{The mass-radius relation for constant density stars for $V=0$, and for three different values of the	constants $\beta_1$ and $\beta_2$: $\beta_1=0.10$  and  $\beta_2=0.15$  (dashed curve),  $\beta_1=0.10$  and  $\beta_2=-0.15$  (dotted curve), and  $\beta_1=-0.20$  and  $\beta_2=0.15$ (dot-dashed curve). The initial values of the vector field are $\bar{h}_0=0.1$ and $\bar{h}^\prime_0=0.5$. The solid curve represents the standard general relativistic  mass-radius relation for constant density stars. }
	\label{consdens-mr-v0}
\end{figure}

The mass-radius relation for this case in shown in Fig.~\ref{consdens-mr-v0}. To obtain the $M-R$ plots we have considered the central pressure in the range $1.89\times10^{32}\, {\rm erg/cm}^3\leq p\leq 2.16\times 10^{35}\,{\rm erg/cm}^3$.

\subsubsection{The case $V=\lambda+a\Lambda_{\mu }\Lambda^{\mu }$}

Next, we consider the effects of a non-zero potential, given by $V=\lambda+a\Lambda_{\mu }\Lambda^{\mu }$, where $a$ and $\lambda$ are constants, on the quark star structure.  The dimensionless parameters in the potential are defined as
\begin{align}
\lambda=\rho_c \bar{\lambda},\quad a=\f{1}{\rho_{c}}\bar{a}.
\end{align}

We also assume that $\lambda$ cannot exceed the central density $\rho_c$, and hence we will take $\bar{\lambda}<1$. Similarly, we will impose the restriction $\bar{a}<1$ on the dimensionless coefficient of $\Lambda_0\Lambda^0$.

\begin{figure*}[htbp]
	\centering
	\includegraphics[width=8.0cm]{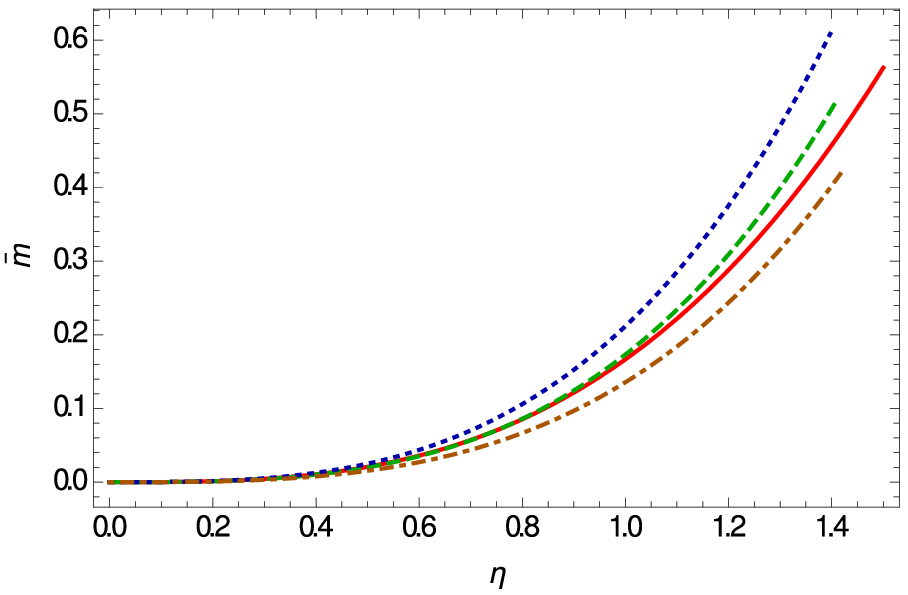}\hspace{.4cm}
	\includegraphics[width=8.0cm]{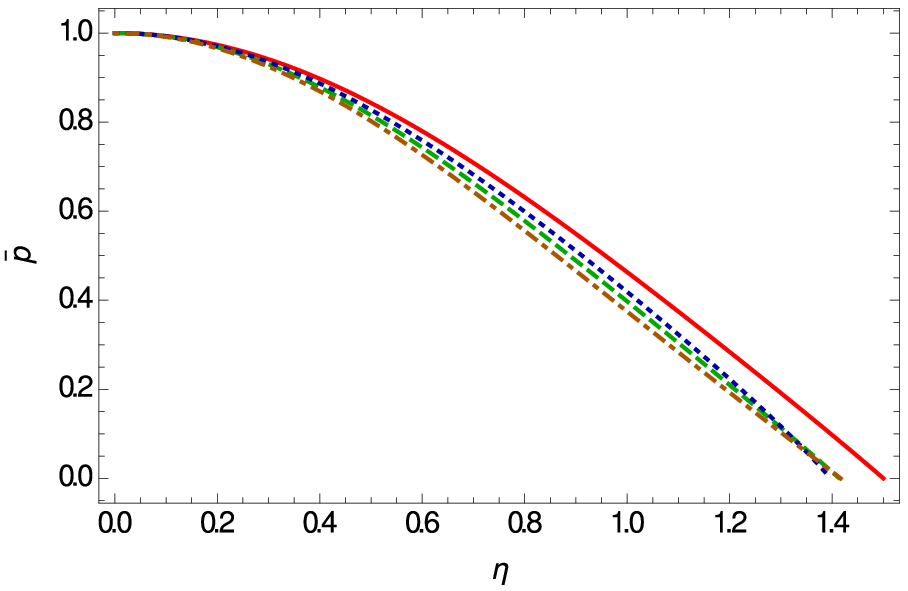}
	\caption{Variation of the interior mass and pressure profiles of constant density star as a function of the radial distance from the center of the  star $\eta$, for $V=\lambda+a\Lambda_{\mu }\Lambda^{\mu }$ and for three different values of the
		constants $\bar{\lambda}$ and $\bar{a}$: $\bar{\lambda}=0$  and  $\bar{a}=-0.5$  (dashed curve),  $\bar{\lambda}=0.21$  and  $\bar{a}=0.0$  (dotted curve), and   $\bar{\lambda}=-0.21$  and  $\bar{a}=0.0$ (dot-dashed curve).  For the central pressure of the star we have adopted the value $p_{c}=1.93\times 10^{35} {\rm erg/cm}^3$, while $\bar{h}_0=0.1$ and $\bar{h}^\prime_0=0.5$. The solid curve represents the standard general relativistic  mass and density profile for constant density stars. }
	\label{consdens-mass-dens-v}
\end{figure*}

\begin{figure}[htbp]
	\centering
	\includegraphics[width=8.0cm]{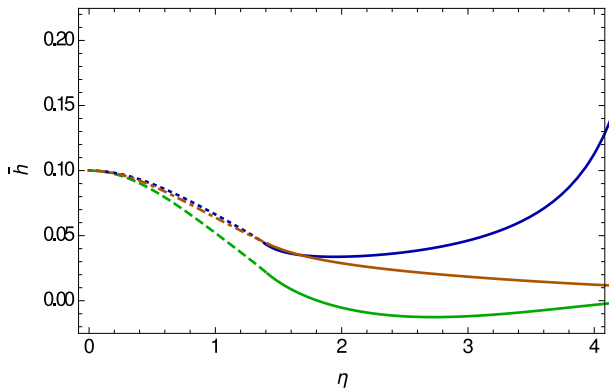}
	\caption{Variation of the temporal component of dark energy vector field inside and outside (solid lines) the constant density star as a function of the radial distance from the center of the  star $\eta$ for $V=\lambda+a\Lambda_{\mu }\Lambda^{\mu }$, and for three different values of the constants $\bar{\lambda}$ and $\bar{a}$: $\bar{\lambda}=0$  and  $\bar{a}=-0.5$  (dashed curve),  $\bar{\lambda}=0.21$  and  $\bar{a}=0.0$  (dotted curve), and   $\bar{\lambda}=-0.21$  and  $\bar{a}=0.0$ (dot-dashed curve).    For the central pressure of the star we have adopted the value $p_{c}=1.93\times 10^{35} {\rm erg/cm}^3$, while $\bar{h}_0=0.1$ and $\bar{h}^\prime_0=0.5$.  }
	\label{consdens-vec-v}
\end{figure}

In the following we set $\beta_1=0.10$ and  $\beta_2=0.15$, and consider the role of the different values of the parameters in the potential term.  
The behaviors of the mass and pressure  inside the constant density star are shown in Figs.~\ref{consdens-mass-dens-v}. One can see the behavior of the non-zero component of the vector dark energy $\bar{h}$, inside and outside of the constant density star in Fig. \ref{consdens-vec-v}. Inside the star it is monotonically decreasing function of $\eta$. However outside the star, the behavior of $\bar{h}$ strongly depends on the potential parameters.

The physical properties of the constant density stars are dependent on the potential parameters, and they determine some significant differences as compared to the standard general relativistic behavior. The dark energy vector field has a similar behavior to the $V=0$ case.

\begin{figure}[htbp]
	\centering
	\includegraphics[width=8.0cm]{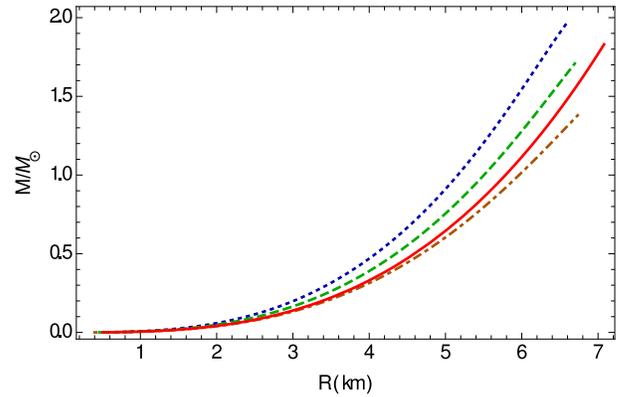}
	\caption{The mass-radius relation for constant density stars for $V=\lambda+a\Lambda_{\mu }\Lambda^{\mu }$ and for three different values of the constants $\bar{\lambda}$ and $\bar{a}$: $\bar{\lambda}=0$  and  $\bar{a}=-0.5$  (dashed curve),  $\bar{\lambda}=0.21$  and  $\bar{a}=0.0$  (dotted curve), and   $\bar{\lambda}=-0.21$  and  $\bar{a}=0.0$ (dot-dashed curve).   The initial values are $\bar{h}_0=0.1$ and $\bar{h}^\prime_0=0.5$. The solid curve represents the standard general relativistic  mass-radius relation for constant density stars. }
	\label{consdens-mr-v}
\end{figure}

The mass-radius relation for this case is presented in Fig.~\ref{consdens-mr-v}, with the central pressure varying in the range $1.89\times 10^{32}\, {\rm erg/cm}^3$ and $ 2.16\times 10^{35}\,{\rm erg/cm}^3$.

Significant differences with respect to the standard general relativistic case can be observed, indicating a dependence of the mass of the Einstein dark energy model on the potential parameters.

\subsection{MIT bag model strange quark stars}

A large number of theoretical and experimental investigations of the baryonic structure indicates that {\it strange quark matter, consisting of the
$u$ (up), $d$ (down) and $s$ (strange) quarks is energetically the most favorable state of baryonic matter}. The possibility of the existence
of stars made of quarks was initially suggested in \cite{Itoh} and \cite{Bo}, and it was later reconsidered in \cite{Wi} and \cite{Hae}.

The theory of quark matter is based on the fundamental Quantum Chromodynamics (QCD) Lagrangian, which is given by \cite{3a,3b,3c}
\bea
L_{QCD}&&=\frac{1}{4}\sum_{a}F_{\mu \nu }^{a}F^{a\mu \nu
}\nonumber\\
&&+\sum_{f=1}^{N_{f}}%
\bar{\psi}\left( i\gamma ^{\mu }\partial _{\mu }- \f12 g\gamma ^{\mu
}A_{\mu }^{a}%
\lambda ^{a}-m_{f}\right) \psi,  \label{s1}
\eea
where the various quark flavors $u$,
$d$, and $s$ are denoted collectively by the subscript $f$, the vector potential $A_{\mu
}^a$  takes values in the Lie algebra with
generators $\lambda ^a$, and $g$ is a  coupling constant, respectively. The gluon field strength
$F_{\mu \nu }^{a}$ is defined as
\begin{equation}
F_{\mu \nu }^{a}=\partial _{\mu }A_{\nu }^{a}-\partial _{\nu
}A_{\mu
}^{a}+gf_{abc}A_{\mu }^{b}A_{\nu }^{c}\,.  \label{s2}
\end{equation}

QCD indicates that at short distances (or high momenta $Q^{2}$) the quark-quark interactions weaken. Moreover,
 for high
momenta $Q^{2}$ the coupling constant $g^{2}\left( Q^{2}\right) $ vanishes, and for $N_f\rightarrow 33/2$ it tends to infinity.

{\it By neglecting the quark
	masses}, and considering that the interactions of quarks and gluons are weak, it follows that the equation of state of strange quark matter
is given by \cite{3a,3b,3c}
\begin{eqnarray}
	\rho &=&\left( 1-\frac{15}{4\pi }\alpha _{s}\right)
	\frac{8\pi ^{2}}{15}{\mathcal T}^{4}+N_{f}\left( 1-\frac{50}{21\pi
	}\alpha_{s}\right) \frac{7\pi ^{2}}{10} {\mathcal T}^{4}
	\nonumber\\
	&&+ \sum_{f}3\left( 1-2\frac{\alpha _{s}}{\pi }\right) \left( \pi
	^{2}{\mathcal T}^{2}+\frac{\mu
		_{f}^{2}}{2}\right) \frac{\mu _{f}^{2}}{\pi ^{2}}+B,  \label{s4}
\end{eqnarray}
where ${\mathcal T}\neq 0$ denotes {\it the temperature of the quark-gluon plasma}, $B$ is the difference between the energy density of the
perturbative and non-perturbative QCD vacuum ({\it the bag constant}), $\mu _f$ is the chemical potential,  $N_f$ is the number of active quark flavors, and $\alpha _s$ is the strong interaction coupling constant, respectively. Equivalently, we obtain
\begin{equation}
	p+B=\sum_{i=u,d,s,e^{-},\mu ^{-}}p_{i}.  \label{s7}
\end{equation}

In its simplest form the Bag Model equation of state takes the form \cite{3a,3b,3c}
\be\label{qeos}
\rho ({\mathcal T})=\sigma _{SB}{\mathcal T}^4+B, \; p({\mathcal T})=\frac{\sigma _{SB}}{3}{\mathcal T}^4-B,
\ee
where the energy density $\rho$ and the pressure $p$ of the quark gluon plasma  are assumed to have a simple dependence on the temperature ${\mathcal T}$,
modified, with respect to the radiation gas, by the addition of the positive constant $B$.  The Stefan-Boltzmann 
constant $\sigma_{SB}$ in Eq.~(\ref{qeos}) is given by $\sigma_{SB}=\left(\pi ^2/30\right)\left(d_B+7d_F/8\right)$, where $d_B$ and $d_F$ denote the degeneracy factors for the gluons, and quarks and antiquarks, respectively  \cite{3a,3b,3c}. Eq.~(\ref{qeos}) is assumed to be valid at temperatures ${\mathcal T} > {\mathcal T}_c$, where ${\mathcal T}_c$ is the critical temperature of a first order phase transition in the $SU(3)$ quark-gluon plasma model, or it is the temperature of a smooth crossover in the
full QCD theory \cite{3c}.

Therefore, from Eq.~(\ref{qeos}) it follows that {\it the thermodynamic parameters of the strange quark-gluon plasma are related, in the first order of approximation, by the MIT bag model equation of state}, given by
\begin{equation}
p=\frac{1}{3}\left( \rho -4B\right),  \label{s6}
\end{equation}

Hence, in the MIT approximate description quarks and gluons can freely move inside the bag, and by squeezing the bags against each other, the deconfined phase can appear.  At high baryon numbers and high energy densities, the components of the quark-gluon plasma can move freely over large regions inside a star. When investigating stellar strange quark-gluon plasma the assumption of the electric charge neutrality must also be imposed,  which can be formulated generally as  
\be
\sum_{i=u,d,s,e^{-},\mu
^{-}}q_{i}n_{i}=0. 
\ee
For a star formed from massless
$u$, $d$ and $s$ quarks, the charge neutrality condition is given by
 $2n_{u}/3=\left(n_{d}+n_{s}\right) /3$.

 Eq.~(\ref{s6}) corresponds to {\it the equation of state of a system of massless particles, with
perturbative interactions}, and negative corrections due to the QCD trace anomaly. For
$\alpha _{s}=0.5$,  the
energy density of a strange quark-gluon plasma is lessened, at a given temperature, by a factor of the order of two \cite{3a,3b,3c}. In the following, we will investigate the properties of dense stellar objects obeying the equation of state (\ref{s6}) in the Einstein dark energy model. For the quark model equation of state the trace of the energy-momentum tensor (or the trace anomaly) is $T=-\rho +3p=-4B<0$.

For the bag constant we adopt the numerical value  $B=1.03\times 10^{14}\, {\rm g/cm}^3$.  For this case, in general relativity, the maximum mass of the star is $M/M_\odot=2.00$ with radius $R=10.92\; {\rm km} $ \cite{Wi, Hae}, and central density $\rho_c=1.98\times10^{15}\, {\rm g/cm}^3$.

We will consider two separate cases corresponding to zero potential, and non-zero potential, respectively, for the dark energy vector. In all cases the central density is  $ \rho_{c}=2.45\times 10^{15} {\rm g/cm}^3$, and the initial conditions for the temporal component of the vector field are $\bar{h}_0=0.1$, and $\bar{h}^\prime_0=0.5$	. The stop point in integration is where the pressure becomes zero, i.e., $\rho=4B$.

\subsubsection{The case $V=0$}

\begin{figure*}[htbp]
	\centering
	\includegraphics[width=8.0cm]{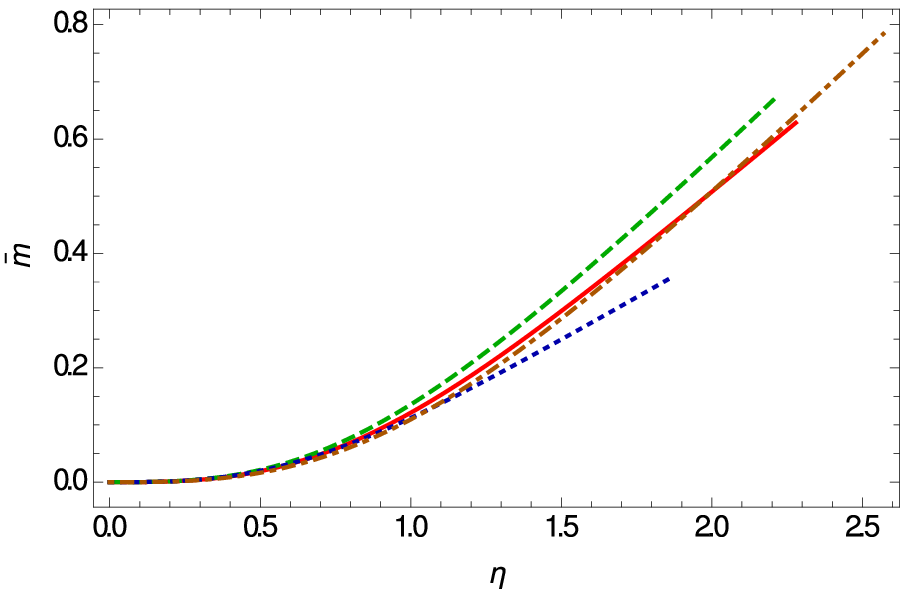}\hspace{.4cm}
	\includegraphics[width=8.0cm]{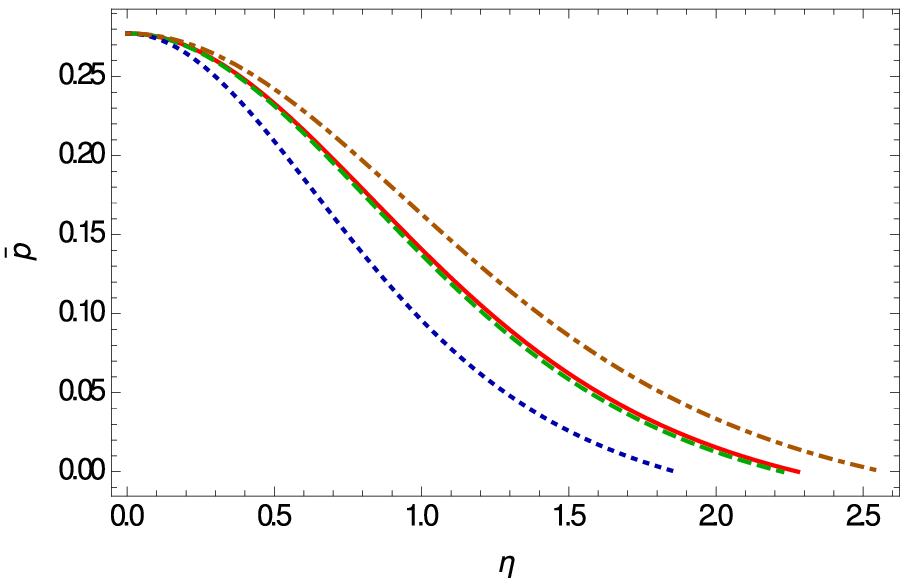}
	\caption{Variation of the interior mass and pressure profiles of MIT quark star as a function of the radial distance from the center of the  star $\eta$ for $V=0$, and for three different values of the
		constants $\beta_1$ and $\beta_2$: $\beta_1=0.10$  and  $\beta_2=0.15$  (dashed curve),  $\beta_1=0.10$  and  $\beta_2=-0.15$  (dotted curve), and  $\beta_1=-0.20$  and  $\beta_2=0.15$ (dot-dashed curve).  For the central density of the star we have adopted the value $\rho_{c}=2.45\times 10^{15} {\rm g/cm}^3$, while $\bar{h}_0=0.1$ and $\bar{h}^\prime_0=0.5$. The solid curve represents the standard general relativistic  mass and density profile for MIT quark stars. }
	\label{MIT-mass-dens-v0}
\end{figure*}

\begin{figure}[htbp]
	\centering
	\includegraphics[width=8.0cm]{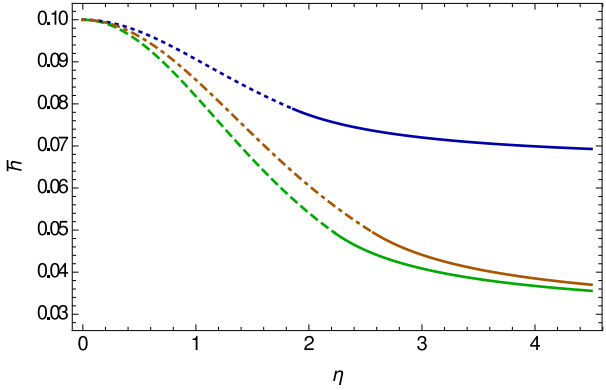}
	\caption{Variation of the temporal component of dark energy vector field inside and outside (solid lines) of the MIT quark star as a function of the radial distance from the center of the  star $\eta$ for $V=0$, and for three different values of the
		constants $\beta_1$ and $\beta_2$: $\beta_1=0.10$  and  $\beta_2=0.15$  (dashed curve),  $\beta_1=0.10$  and  $\beta_2=-0.15$  (dotted curve), and  $\beta_1=-0.20$  and  $\beta_2=0.15$ (dot-dashed curve).  For the central density of the star we have adopted the value $\rho_{c}=2.45\times 10^{15} {\rm g/cm}^3$, while $\bar{h}_0=0.1$ and $\bar{h}^\prime_0=0.5$.  }
	\label{MIT-vec-v0}
\end{figure}

The behaviors of mass and pressure inside the MIT quark star are presented in Fig.~\ref{MIT-mass-dens-v0}, and Fig. \ref{MIT-vec-v0} shows the  behavior of  non-zero component of the vector dark  energy $\bar{h}$ inside and outside of the MIT quark star. As one can see from the figures, the mass distribution is a monotonically increasing function of $r$. On the other hand, the pressure is monotonically decreasing, and becomes zero for a finite value $r=R$ of the radial coordinate, with $R$ giving the radius of the star. The variation of the temporal component of the Einstein dark energy vector field is shown in Fig.~\ref{MIT-vec-v0}. $\bar{h}$ reaches its maximum at the stellar center, and decreases toward the star's surface. Finally, it reaches a constant value outside the star.

\begin{figure}[htbp]
	\centering
	\includegraphics[width=8.0cm]{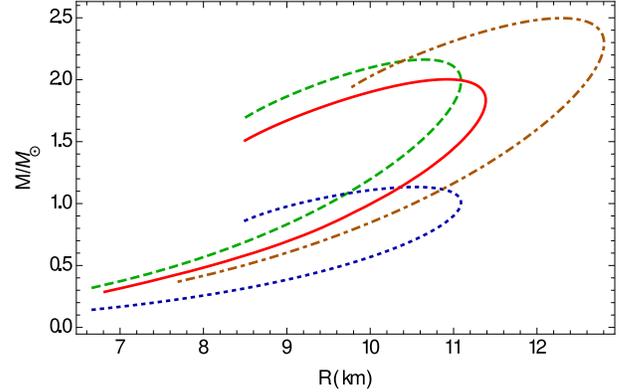}
	\caption{The mass-radius relation for MIT quark stars for $V=0$ and for three different values of the	constants $\beta_1$ and $\beta_2$: $\beta_1=0.10$  and  $\beta_2=0.15$  (dashed curve),  $\beta_1=0.10$  and  $\beta_2=-0.15$  (dotted curve), and  $\beta_1=-0.20$  and  $\beta_2=0.15$ (dot-dashed curve). The initial values are $\bar{h}_0=0.1$ and $\bar{h}^\prime_0=0.5$. The solid curve represents the standard general relativistic  mass-radius relation for MIT quark stars. }
	\label{MIT-mr-v0}
\end{figure}

The mass-radius relation for this case is presented in Fig.~\ref{MIT-mr-v0}. To obtain Fig.~\ref{MIT-mr-v0} we have considered a range of central densities varying between $4.6\times 10^{14}\, {\rm g/cm}^3 $, and $2\times 10^{16}\, {\rm g/cm}^3 $. Different values of the coupling parameters can induce large departures as compared to the standard general relativistic case, leading to stars having both much larger, and much smaller, maximum masses.

Some specific parameters of the MIT bag model quark stars are presented in Table~\ref{MIT-v0-tab}. For a negative value of the parameter $\beta _1$, a significant increase in the maximum mass of the quark does occur. On the other hand a negative $\beta _2$ leads to a drastic decrease in the maximum mass, with values of the order of $1.13M_{\odot}$. The maximum mass values are reached for central densities of the order of $\rho _c\approx 2\times 10^{15}{\;}{\rm g/cm}^3$.

\begin{table}[h!]
\begin{center}
	\begin{tabular}{|c|c|c|c|}
		\hline
		$\beta_1$ &~~~$-0.20$~~~&$~~~0.10~~~~$&$~~~0.10~~~~$ \\
		\hline
		$\beta_2$ &~~~$0.15$~~~&$~~~-0.15~~~~$&$~~~0.15~~~~$ \\
		\hline
		\quad$M_{max}/M_{\odot}$\quad& $~~~2.49~~~$& $~~~1.13~~~$& $~~~2.16~~~$\\
		\hline
		$~~~R\,({\rm km})~~~$& $~~~12.31~~~$& $~~~10.52~~~$& $~~~10.66~~~$\\
		\hline
		$~~~\rho_{c} \times 10^{-15}\,({\rm g/cm}^3)~~~$& $~~~1.96~~~$& $~~~2.22~~~$& $~~~1.96~~~$\\
		\hline
	\end{tabular}
	\caption{The maximum masses and corresponding radii and central densities for the MIT bag model stars for $V=0$.}\label{MIT-v0-tab}
\end{center}
\end{table}

\subsubsection{The case $V=\lambda+a\Lambda_{\mu }\Lambda^{\mu }$}

\begin{figure*}[htbp]
	\centering
	\includegraphics[width=8.0cm]{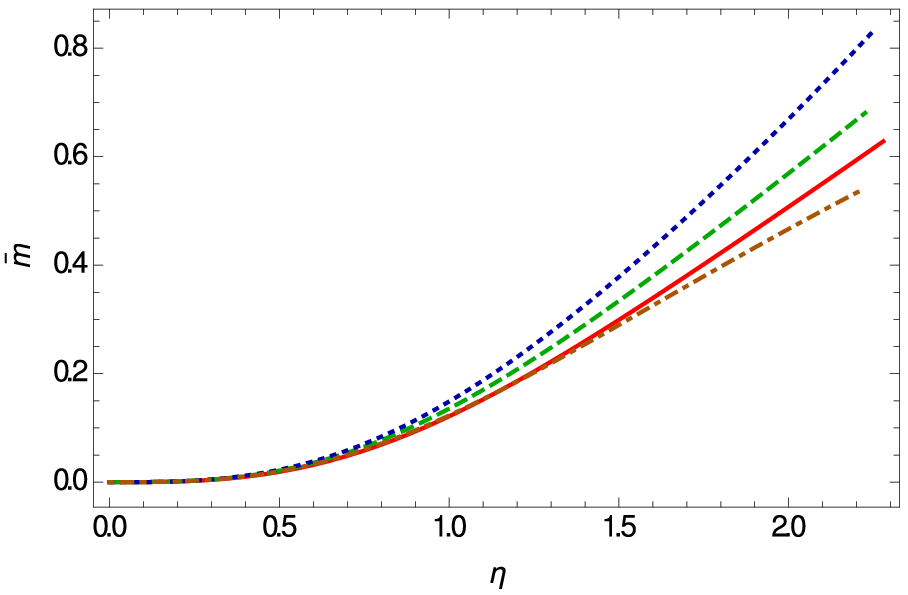}\hspace{.4cm}
	\includegraphics[width=8.0cm]{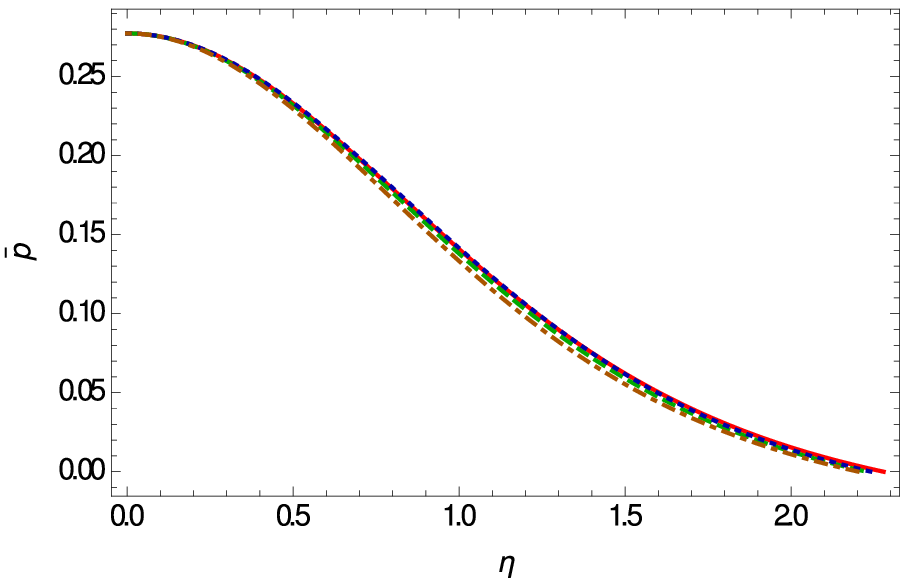}
	\caption{Interior mass and pressure profiles of MIT quark stars as a function of the dimensionless radial distance from the center of the  star $\eta$ for  $V=\lambda+a\Lambda_{\mu }\Lambda^{\mu }$, and for three different values of the constants $\bar{\lambda}$ and $\bar{a}$: $\bar{\lambda}=0$  and  $\bar{a}=-0.5$  (dashed curve),  $\bar{\lambda}=0.06$  and  $\bar{a}=0.0$  (dotted curve), and   $\bar{\lambda}=-0.06$  and  $\bar{a}=0.0$ (dot-dashed curve).  For the central density of the star we have adopted the value $\rho_{c}=2.45\times 10^{15} {\rm g/cm}^3$, while $\bar{h}_0=0.1$ and $\bar{h}^\prime_0=0.5$. The solid curve represents the standard general relativistic  mass and density profile for MIT quark stars. }
	\label{MIT-mass-dens-v}
\end{figure*}

\begin{figure}[htbp]
	\centering
	\includegraphics[width=8.0cm]{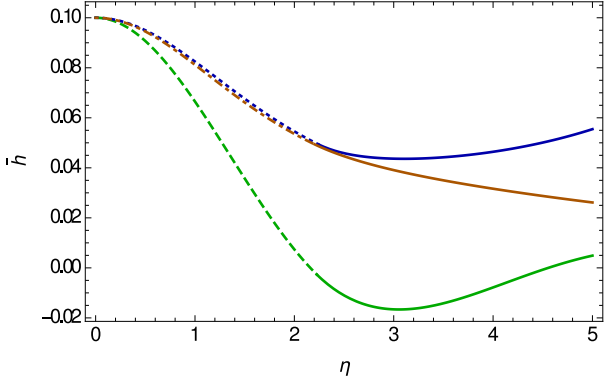}
	\caption{Variation of the temporal component of dark energy vector field inside and outside (solid lines) of the MIT quark star as a function of the radial distance from the center of the  star $\eta$ for  $V=\lambda+a\Lambda_{\mu }\Lambda^{\mu }$, and for three different values of the constants $\bar{\lambda}$ and $\bar{a}$: $\bar{\lambda}=0$  and  $\bar{a}=-0.5$  (dashed curve),  $\bar{\lambda}=0.06$  and  $\bar{a}=0.0$  (dotted curve), and   $\bar{\lambda}=-0.06$  and  $\bar{a}=0.0$ (dot-dashed curve).    For the central density of the star we have adopted the value $\rho_{c}=2.45\times 10^{15} {\rm g/cm}^3$, while $\bar{h}_0=0.1$ and $\bar{h}^\prime_0=0.5$.  }
	\label{MIT-vec-v}
\end{figure}

Now,  the effects of a non-zero potential, given by $V=\lambda+a\Lambda_{\mu }\Lambda^{\mu }$, are considered on the quark star structure.
 In the following we have set $\beta_1=0.10$ and  $\beta_2=0.15$, and we investigate the role of the different values of the parameters in the potential term. The behaviors of the mass and pressure inside the MIT quark star are presented in Fig.~\ref{MIT-mass-dens-v}.  The mass is a monotonically increasing function inside star, while the pressure becomes zero on the stellar surface. The evolution of $\bar{p}$ does not show a significant dependence on the potential parameters. One can also see the behavior of the non-zero component of the vector dark  energy $\bar{h}$ inside and outside the MIT quark star in Fig. \ref{MIT-vec-v}.   The vector field monotonically decreases inside the star. The evolution of $\bar{h}$ outside the star depends on the parameters of the potential.

\begin{figure}[htbp]
	\centering
	\includegraphics[width=8.0cm]{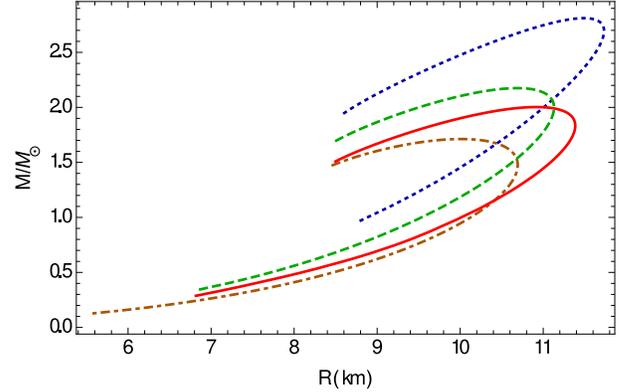}
	\caption{The mass-radius relation for MIT quark stars for  $V=\lambda+a\Lambda_{\mu }\Lambda^{\mu }$ and for three different values of the constants $\bar{\lambda}$ and $\bar{a}$: $\bar{\lambda}=0$  and  $\bar{a}=-0.5$  (dashed curve),  $\bar{\lambda}=0.06$  and  $\bar{a}=0.0$  (dotted curve), and   $\bar{\lambda}=-0.06$  and  $\bar{a}=0.0$ (dot-dashed curve).   The initial values are $\bar{h}_0=0.1$ and $\bar{h}^\prime_0=0.5$. The solid curve represents the standard general relativistic  mass-radius relation for MIT quark stars. }
	\label{MIT-mr-v}
\end{figure}

The mass-radius relation for this case in shown in Fig.~\ref{MIT-mr-v}. The presence of the potential of the vector field strongly influences the maximum mass of the star, which can reach values as high as $2.81M_{\odot}$. Selected maximum mass values of the quark stars in the Einstein dark energy model are presented in Table~\ref{MIT-v-tab}.

\begin{table}[h!]
	\begin{center}
		\begin{tabular}{|c|c|c|c|}
			\hline
			$\bar{\lambda}$ &~~~$-0.06$~~~&$~~~0.0~~~~$&$~~~0.06~~~~$ \\
			\hline
			$\bar{a}$ &~~~$0.0$~~~&$~~~-0.5~~~~$&$~~~0.0~~~~$ \\
			\hline
			\quad$M_{max}/M_{\odot}$\quad& $~~~1.71~~~$& $~~~2.17~~~$& $~~~2.81~~~$\\
			\hline
			$~~~R\,({\rm km})~~~$& $~~~10.01~~~$& $~~~10.66~~~$& $~~~11.49~~~$\\
			\hline
			$~~~\rho_{c} \times 10^{-15}\,({\rm g/cm}^3)~~~$& $~~~3.24~~~$& $~~~1.96~~~$& $~~~1.13~~~$\\
			\hline
		\end{tabular}
		\caption{The maximum masses and the corresponding radii  and central densities for the MIT bag model stars for $V=\lambda+a\Lambda_{\mu }\Lambda^{\mu }$, and for $\beta_1=0.10$ and  $\beta_2=0.15$  .}\label{MIT-v-tab}
	\end{center}
\end{table}

\subsection{CFL quark stars}

At very high densities the quarks may form Cooper pairs, whose color properties are correlated with their flavor properties in a one-to-one correspondence between three color pairs and three flavor pairs. The CFL phase is the highest-density phase of three-flavor colored matter. There is presently an almost general consensus that, even if the quark masses are unequal, the CFL  phase of quarks represents the ground state of matter, at least for asymptotic densities \cite{Al1,cfl2, Al2,Al3,Al4, Zoltan}. The equal number of flavors is imposed by symmetry considerations. In this phase there is no need for electrons to be present, since the quark-gluon plasma is neutral. The properties of the quark matter in the  CFL phase depend significantly on the values of the deconfinement phase transition density, and on the CFL gap parameter. From both theoretical and experimental points of view the numerical values of these parameters are not well known.  For quark matter in the CFL phase the free energy density $\Omega _{CFL}$ is given by \cite{42}
\bea
\Omega _{CFL}\left(\mu ,
\mu_e\right)&=&\Omega_{CFL}^{quarks}\left(\mu \right)+\Omega
_{CFL}^{GB}\left(\mu ,\mu _e \right)\nonumber\\
&&+\Omega _{CFL}^{electrons}\left(\mu _e\right),
\eea
where by $\Omega _{CFL}^{GB}$ we have denoted the contribution from the Goldstone
bosons, coming from the breaking of chiral symmetry of the CFL
phase, and $\mu _e$ is the chemical potential of the electrons. {\it If the mass} $m_{s}$ {\it of the} $s$ {\it quark is of the same order of magnitude
as the chemical potential} $\mu $, one can approximate the
thermodynamical potential of the quark matter in the CFL phase as \cite{LuHo02}
\bea
\Omega _{CFL}&=&-\frac{3\mu ^{4}}{4\pi ^{2}}+\frac{3m_{s}^{2}}
{4\pi ^{2}}-%
\frac{1-12\ln \left( m_{s}/2\mu \right) }{32\pi ^{2}}m_{s}^{4}
    \nonumber\\
&&-\frac{3}{\pi
^{2}}\Delta ^{2}\mu ^{2}+B,
\eea
where $\Delta $ denotes the gap energy. From $\Omega _{CFL}$ the expression
for the pressure $P$ of the quark matter in the CFL phase is
obtained as a function of the energy density $\rho\overline{} $ as \cite{LuHo02}
\begin{equation}\label{pres}
P=\frac{1}{3}\left( \rho -4B\right) +\frac{2\Delta
^{2}\delta ^{2}}{\pi
^{2}}-\frac{m_{s}^{2}\delta ^{2}}{2\pi ^{2}},
\end{equation}
where
\begin{equation}
\delta ^{2}=-\alpha +\sqrt{\alpha ^{2}+\frac{4}{9}\pi ^{2}\left(
\varepsilon
-B\right) },
\end{equation}%
and
\be\label{alpha}
\alpha =-\frac{m_{s}^{2}}{6}+\frac{2\Delta ^{2}}{3}.
\ee

Hence, Eq.~(\ref{pres}) gives finally the equation of state of the quark matter in the CFL phase as
\begin{equation}\label{pres2}
P=\frac{1}{3}\left(\rho -4B\right) +\frac{3 \alpha \delta
^{2}}{\pi^{2}},
\end{equation}
which will be used in the analysis of the Einstein dark energy stars outlined below.

We consider the cases with $\Delta=300\,{\rm MeV}$, $m_s=150\,{\rm MeV}$ and $B=1.15\times 10^{14}\,{\rm g/cm}^3$. The range of central densities is between $4.7\times 10^{14}\, {\rm g/cm}^3 $ and $6.8\times 10^{15}\, {\rm g/cm}^3 $. In the standard general relativistic case, the maximum mass of the CFL quark star is $M/M_\odot=2.10$, with the radius $R=11.21\, {\rm km} $, and central density $\rho_c=1.88\times10^{15}\, {\rm g/cm}^3$.

In the following we consider again two separate cases corresponding to a zero potential, and a non-zero potential, respectively, for the vector dark energy. In all cases the central density is  $ \rho_{c}=2.45\times 10^{15} {\rm g/cm}^3$,and the initial conditions for the temporal component of the vector field are $\bar{h}_0=0.1$ and $\bar{h}^\prime_0=0.5$, respectively. The stop point in integration is where the pressure becomes , i.e. $p=0$.

\subsubsection{The case $V=0$}

\begin{figure*}[htbp]
	\centering
	\includegraphics[width=8.0cm]{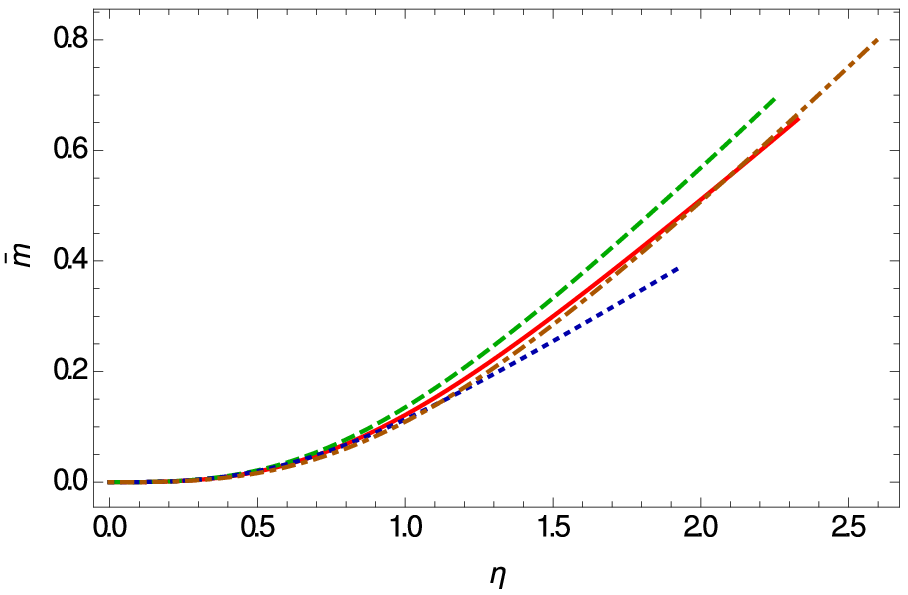}\hspace{.4cm}
	\includegraphics[width=8.0cm]{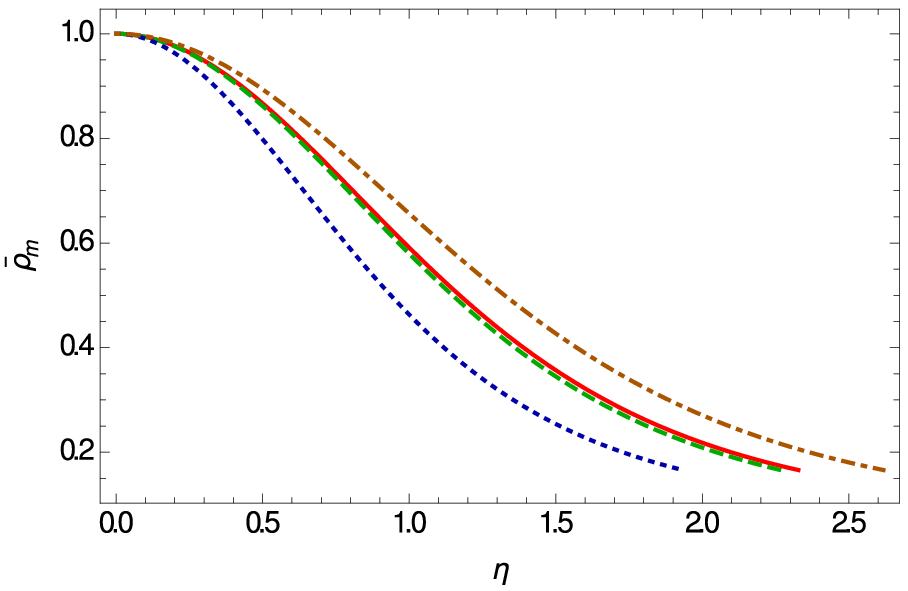}
	\caption{Variation of the interior mass and density profiles of CFL quark stars for $V=0$ as a function of the radial distance from the center of the  star $\eta$, for three different values of the
		constants $\beta_1$ and $\beta_2$: $\beta_1=0.10$  and  $\beta_2=0.15$  (dashed curve),  $\beta_1=0.10$  and  $\beta_2=-0.15$  (dotted curve), and  $\beta_1=-0.20$  and  $\beta_2=0.15$ (dot-dashed curve).  For the central density of the star we have adopted the value $\rho_{c}=2.45\times 10^{15} {\rm g/cm}^3$, while $\bar{h}_0=0.1$ and $\bar{h}^\prime_0=0.5$. The solid curve represents the standard general relativistic  mass and density profile for CFL quark stars. }
	\label{CFL-mass-dens-v0}
\end{figure*}

\begin{figure}[htbp]
	\centering
	\includegraphics[width=8.0cm]{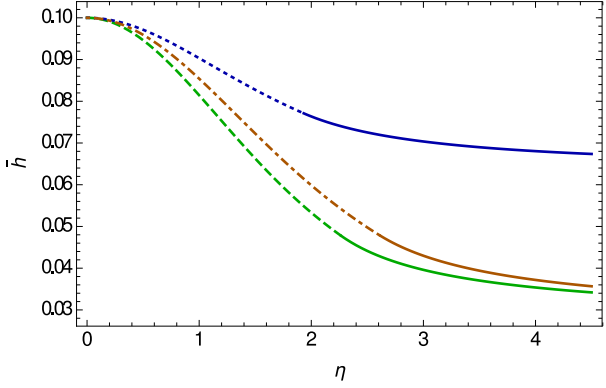}
	\caption{Variation of the temporal component of dark energy vector field inside and outside (solid lines) of the CFL quark star as a function of the radial distance from the center of the  star $\eta$, for three different values of the
		constants $\beta_1$ and $\beta_2$: $\beta_1=0.10$  and  $\beta_2=0.15$  (dashed curve),  $\beta_1=0.10$  and  $\beta_2=-0.15$  (dotted curve), and  $\beta_1=-0.20$  and  $\beta_2=0.15$ (dot-dashed curve).  For the central density of the star we have adopted the value $\rho_{c}=2.45\times 10^{15} {\rm g/cm}^3$, while $\bar{h}_0=0.1$ and $\bar{h}^\prime_0=0.5$.  }
	\label{CFL-vec-v0}
\end{figure}

We first consider the case of a vanishing potential, with $V=0$. The behaviors of the mass and the pressure inside the CFL quark star are depicted in Fig.~\ref{CFL-mass-dens-v0}. The mass and the pressure behave physically inside the star.  The variation of non-zero component of the vector dark  energy $\bar{h}$  in terms of $\eta$ is depicted in Fig. \ref{CFL-vec-v0}.  $\bar{h}$ is a monotonically decreasing function of $\eta$ and asymptotically reaches a constant value outside the star.

\begin{figure}[htbp]
	\centering
	\includegraphics[width=8.0cm]{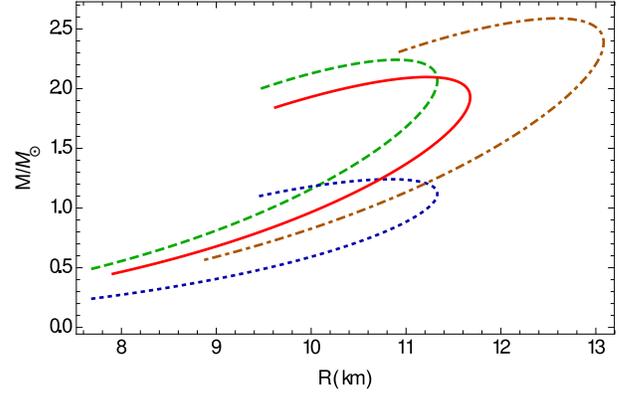}
	\caption{The mass-radius relation for CFL quark stars with $V=0$ for three different values of the	constants $\beta_1$ and $\beta_2$: $\beta_1=0.10$  and  $\beta_2=0.15$  (dashed curve),  $\beta_1=0.10$  and  $\beta_2=-0.15$  (dotted curve), and  $\beta_1=-0.20$  and  $\beta_2=0.15$ (dot-dashed curve). The initial values are $\bar{h}_0=0.1$ and $\bar{h}^\prime_0=0.5$. The solid curve represents the standard general relativistic  mass-radius relation for CFL quark stars. }
	\label{CFL-mr-v0}
\end{figure}

The mass-radius relation for this case, shown in Fig.~\ref{CFL-mr-v0}, indicates again significant departures from the standard relativistic case, with the maximum masses of the CFL quark stars strongly dependent on the model parameters.

A few selected values of the maximum masses and radii of the CFL quark stars are presented in Table~\ref{CFL-v0-tab}. The change in the numerical values of the coupling constants $\beta _1$ and $\beta _2$ leads to significant variations in the numerical values of the maximal masses of the stars, which can have both higher and lower values as compared to the standard general relativistic case.

\begin{table}[h!]
	\begin{center}
		\begin{tabular}{|c|c|c|c|}
			\hline
			$\beta_1$ &~~~$-0.20$~~~&$~~~0.10~~~~$&$~~~0.10~~~~$ \\
			\hline
			$\beta_2$ &~~~$0.15$~~~&$~~~-0.15~~~~$&$~~~0.15~~~~$ \\
			\hline
			\quad$M_{max}/M_{\odot}$\quad& $~~~2.58~~~$& $~~~1.24~~~$& $~~~2.24~~~$\\
			\hline
			$~~~R\,({\rm km})~~~$& $~~~12.58~~~$& $~~~10.83~~~$& $~~~10.89~~~$\\
			\hline
			$~~~\rho_{c} \times 10^{-15}\,({\rm g/cm}^3)~~~$& $~~~1.91~~~$& $~~~2.01~~~$& $~~~1.91~~~$\\
			\hline
		\end{tabular}
		\caption{The maximum masses, and the corresponding radii and central densities for CFL quark star model with $V=0$.}\label{CFL-v0-tab}
	\end{center}
\end{table}

\subsubsection{The case $V=\lambda+a\Lambda_{\mu }\Lambda^{\mu }$}

\begin{figure*}[htbp]
	\centering
	\includegraphics[width=8.0cm]{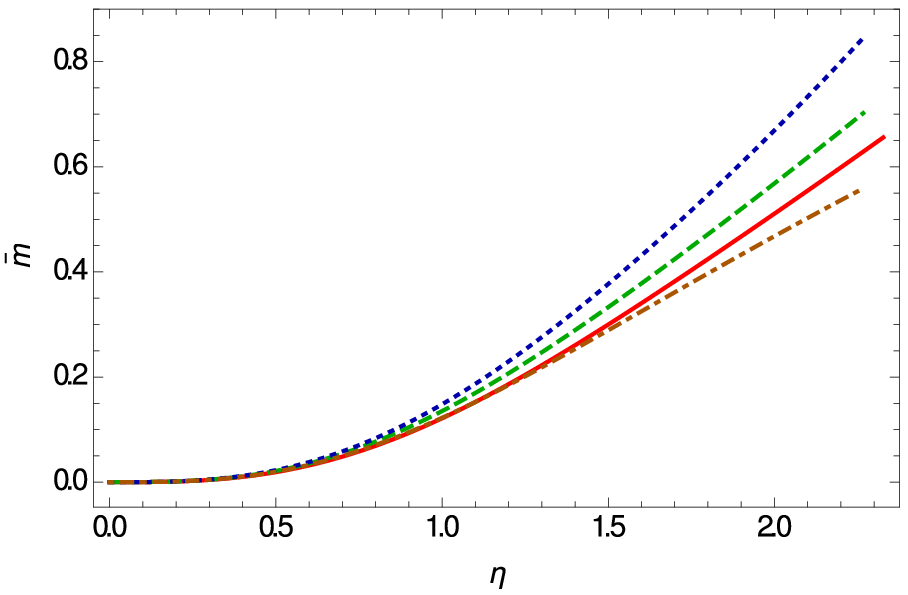}\hspace{.4cm}
	\includegraphics[width=8.0cm]{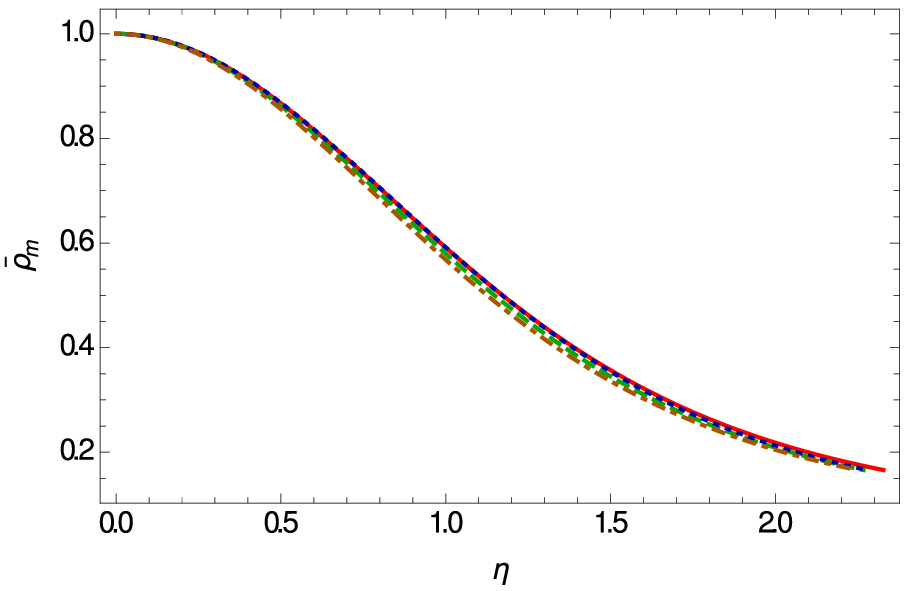}
	\caption{Variation of the interior mass and density profiles of the CFL quark star for $V=\lambda+a\Lambda_{\mu }\Lambda^{\mu }$ as a function of the radial distance from the center of the  star $\eta$, for three different values of the
		constants $\bar{\lambda}$ and $\bar{a}$: $\bar{\lambda}=0$  and  $\bar{a}=-0.5$  (dashed curve),  $\bar{\lambda}=0.06$  and  $\bar{a}=0.0$  (dotted curve), and   $\bar{\lambda}=-0.06$  and  $\bar{a}=0.0$ (dot-dashed curve).  For the central density of the star we have adopted the value $\rho_{c}=2.45\times 10^{15} {\rm g/cm}^3$, while $\bar{h}_0=0.1$ and $\bar{h}^\prime_0=0.5$. The solid curve represents the standard general relativistic  mass and density profile for CFL quark stars. }
	\label{CFL-mass-dens-v}
\end{figure*}

\begin{figure}[htbp]
	\centering
	\includegraphics[width=8.0cm]{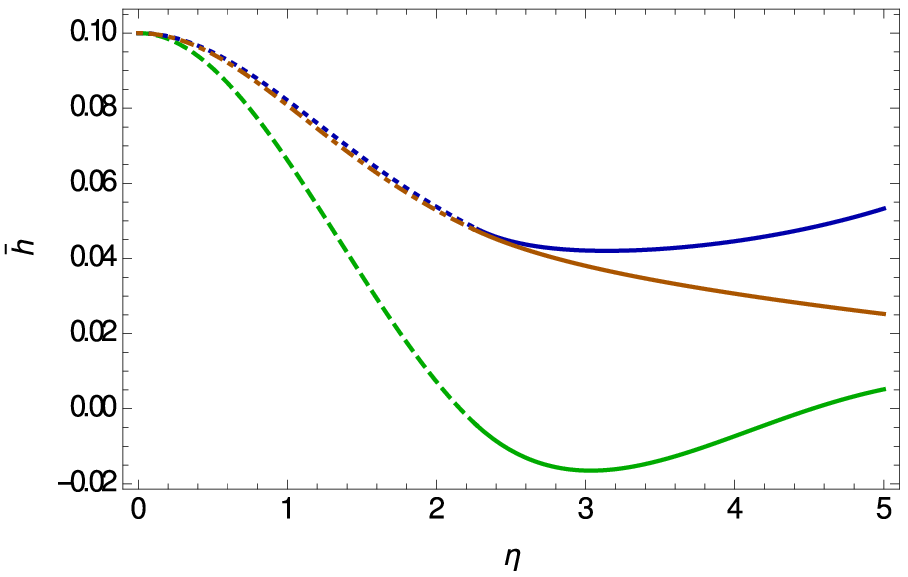}
	\caption{Variation of the temporal component of dark energy vector field inside and outside (solid lines) of the CFL quark star for $V=\lambda+a\Lambda_{\mu }\Lambda^{\mu }$  as a function of the radial distance from the center of the  star $\eta$, for three different values of the constants $\bar{\lambda}$ and $\bar{a}$: $\bar{\lambda}=0$  and  $\bar{a}=-0.5$  (dashed curve),  $\bar{\lambda}=0.06$  and  $\bar{a}=0.0$  (dotted curve), and   $\bar{\lambda}=-0.06$  and  $\bar{a}=0.0$ (dot-dashed curve).    For the central density of the star we have adopted the value $\rho_{c}=2.45\times 10^{15} {\rm g/cm}^3$, while $\bar{h}_0=0.1$ and $\bar{h}^\prime_0=0.5$.  }
	\label{CFL-vec-v}
\end{figure}

Next, we investigate CFL quark stars in the presence of a nonzero potential of the dark vector field, with $V=\lambda+a\Lambda_{\mu }\Lambda^{\mu }$. We fix $\beta_1=0.10$ and  $\beta_2=0.15$, and consider the role of the different values of the parameters in the potential term. 
The behaviors of the mass and  pressure inside the CFL quark star with $V\neq 0$ are depicted in Fig.~\ref{CFL-mass-dens-v}.  The behavior of the non-zero component of the vector dark  energy $\bar{h}$ is shown in Fig. \ref{CFL-vec-v}, inside and outside of the CFL quark star. All these quantities have a physical behavior, similar to the previous cases.

\begin{figure}[htbp]
	\centering
	\includegraphics[width=8.0cm]{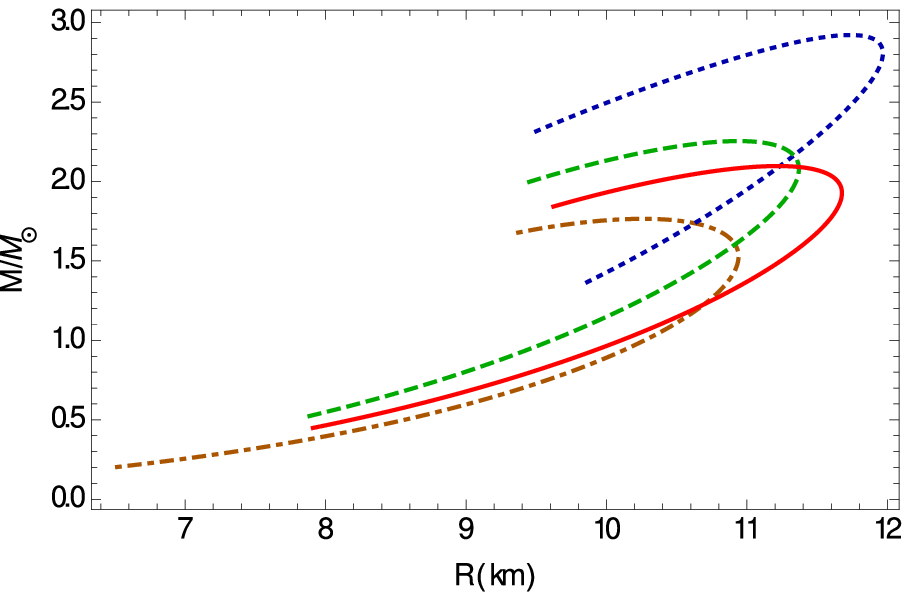}
	\caption{The mass-radius relation for CFL quark stars for $V=\lambda+a\Lambda_{\mu }\Lambda^{\mu }$, and for three different values of the constants $\bar{\lambda}$ and $\bar{a}$: $\bar{\lambda}=0$  and  $\bar{a}=-0.5$  (dashed curve),  $\bar{\lambda}=0.06$  and  $\bar{a}=0.0$  (dotted curve), and   $\bar{\lambda}=-0.06$  and  $\bar{a}=0.0$ (dot-dashed curve).   The initial values are $\bar{h}_0=0.1$ and $\bar{h}^\prime_0=0.5$. The solid curve represents the standard general relativistic  mass-radius relation for CFL quark stars. }
	\label{CFL-mr-v}
\end{figure}

The mass-radius relation for CFL quark stars with $V=\lambda+a\Lambda_{\mu }\Lambda^{\mu }$ is shown in Fig.~\ref{CFL-mr-v}. The main results of the figures are summarized in Table~\ref{CFL-v-tab}, which indicates a significant increase in the maximum mass of the stars, which can reach values as high as $3M_{\odot}$.

\begin{table}[h!]
	\begin{center}
		\begin{tabular}{|c|c|c|c|}
			\hline
			$\bar{\lambda}$ &~~~$-0.06$~~~&$~~~0.0~~~~$&$~~~0.06~~~~$ \\
			\hline
			$\bar{a}$ &~~~$0.0$~~~&$~~~-0.5~~~~$&$~~~0.0~~~~$ \\
			\hline
			\quad$M_{max}/M_{\odot}$\quad& $~~~1.76~~~$& $~~~2.25~~~$& $~~~2.92~~~$\\
			\hline
			$~~~R\,({\rm km})~~~$& $~~~10.26~~~$& $~~~10.92~~~$& $~~~11.73~~~$\\
			\hline
			$~~~\rho_{c} \times 10^{-15}\,({\rm g/cm}^3)~~~$& $~~~3.12~~~$& $~~~1.86~~~$& $~~~1.08~~~$\\
			\hline
		\end{tabular}
		\caption{The maximum mass and corresponding radius and central density for CFL quark stars with $V=\lambda+a\Lambda_{\mu }\Lambda^{\mu }$ for $\beta_1=0.10$ and  $\beta_2=0.15$  .}\label{CFL-v-tab}
	\end{center}
\end{table}

\subsection{Bose-Einstein condensate stars}

Bose-Einstein condensation is assumed to play a key role in many nuclear and quark matter processes. Presently, it is assumed that at ultra-high
densities nuclear matter is formed from a degenerate Fermi gas of quarks. In this system Cooper pairs of quarks
form near the Fermi surface. Therefore, high density nuclear matter in the quark phase can be described,  from
a physical point of view, as a color superconductor \cite{13a,13b}. For strong enough attractive interactions between fermions, and with the temperature dropping below the critical value, the fermions undergo a phase transition into the bosonic
zero mode, and form a Bose-Einstein condensate of quarks \cite{14a,14b,14c}. Hence, in order to obtain a Bose Einstein condensate of fermions one must first form
a BCS state, which can be realized physically under the assumption that the attractive interaction between particles is weak. Hence, high density nuclear matter may exist in a superfluid phase, characterized for single particle excitations by
the existence of an energy gap. The energy gap is formed through the creation of the
Cooper pairs. A Bose Einstein condensate can also be formed when
the attractive interaction between fermions is extremely
strong, leading  to the formation of bound particles (bosons). At the critical temperature $T_c$ the bosons begin to condense into the bosonic zero mode. On the other hand,
the BCS and the BEC states are smoothly connected (crossover), and no phase transition occurs in the system.

The equation of state of the standard Bose-Einstein
condensate with quartic non-linearity is given by a polytropic equation of state  with index $n = 1$, given by
\begin{align}
p=k\rho^2,
\end{align}
where $k$ is a constant.

In the following we consider the cases with $\bar{k}=\rho_c k=0.4$. The range of central density is between $3.96\times 10^{14}\, {\rm g/cm}^3 $ and $7.35\times 10^{15}\, {\rm g/cm}^3 $. For this case, the maximum mass of the standard general relativistic star is $M/M_\odot=2.00$, with radius $R=11.17\, {\rm km} $, and central density $\rho_c=2.58\times10^{15}\, {\rm g/cm}^3$. In the following we consider two separate cases, corresponding to a zero potential, and a non-zero potential, respectively,  for the vector dark energy. In all cases the central density is $ \rho_{c}=2.45\times 10^{15} {\rm g/cm}^3$,and the initial condition for the temporal component of the vector field is $\bar{h}_0=0.1$ and $\bar{h}^\prime_0=0.5$, respectively. The stop point in integration is where $\rho=1.96\times 10^{13}\, {\rm g/cm}^3$.

\subsubsection{The case $V=0$}

\begin{figure*}[htbp]
	\centering
	\includegraphics[width=8.0cm]{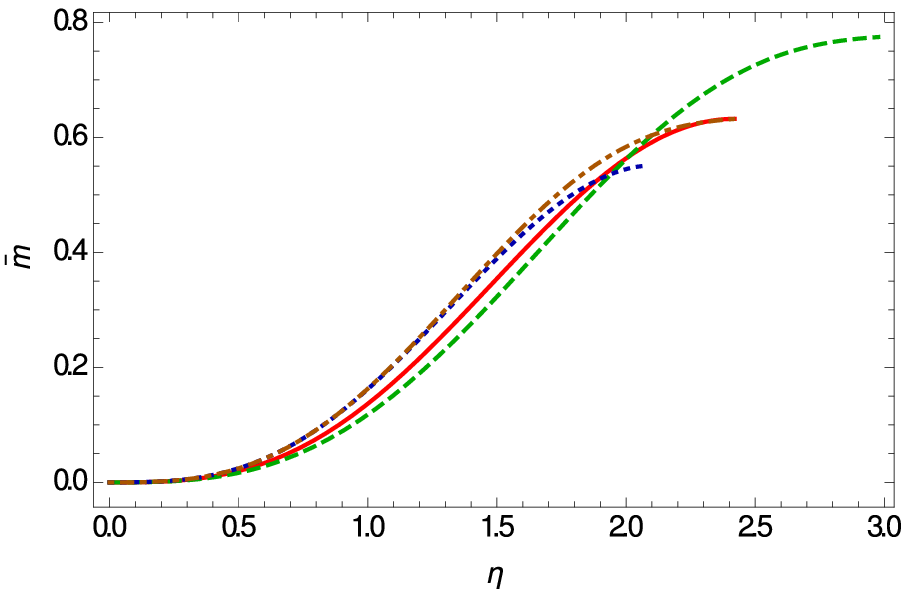}\hspace{.4cm}
	\includegraphics[width=8.0cm]{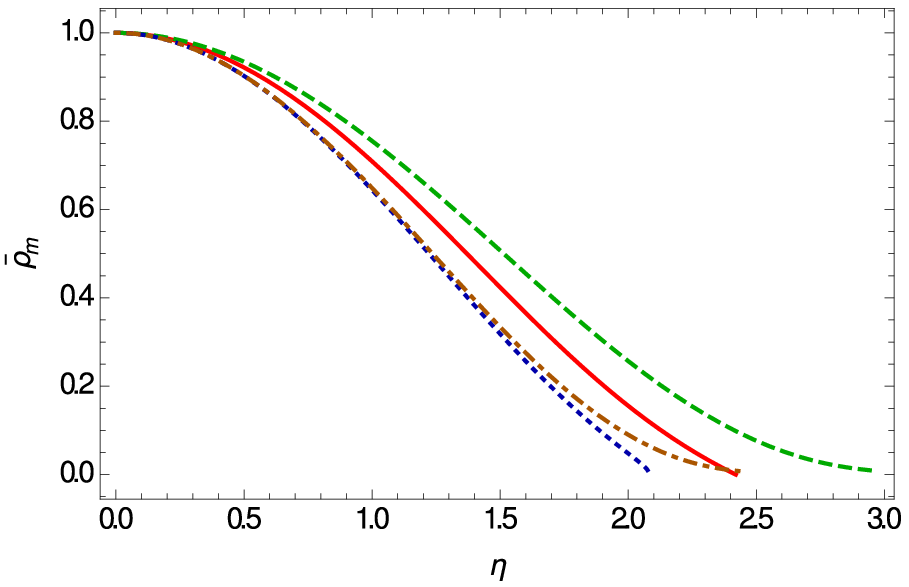}
	\caption{Variation of the interior mass and density profiles of BEC star with $V=0$  as a function of the radial distance from the center of the  star $\eta$, for three different values of the
		constants $\beta_1$ and $\beta_2$: $\beta_1=-0.20$  and  $\beta_2=0.05$  (dashed curve),  $\beta_1=0.20$  and  $\beta_2=-0.01$  (dotted curve), and  $\beta_1=0.20$  and  $\beta_2=0.05$ (dot-dashed curve), respectively.  For the central density of the star we have adopted the value $\rho_{c}=2.45\times 10^{15} {\rm g/cm}^3$, while $\bar{h}_0=0.1$ and $\bar{h}^\prime_0=0.5$, respectively. The stop-point in integration is $\rho=1.96\time10^{13}\, {\rm g/cm}^3$. The solid curve represents the standard general relativistic  mass and density profile for BE stars. }
	\label{BE-mass-dens-v0}
\end{figure*}

\begin{figure}[htbp]
	\centering
	\includegraphics[width=8.0cm]{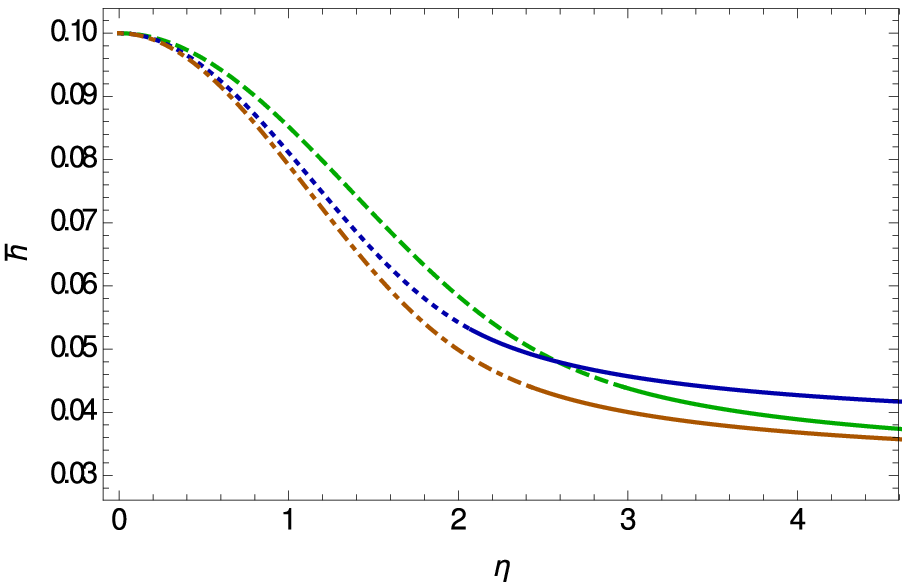}
	\caption{Variation of the temporal component of dark energy vector field inside and outside (solid lines) of the BEC star with $V=0$ as a function of the radial distance from the center of the  star $\eta$, for three different values of the
		constants $\beta_1$ and $\beta_2$:  $\beta_1=-0.20$  and  $\beta_2=0.05$  (dashed curve),  $\beta_1=0.20$  and  $\beta_2=-0.01$  (dotted curve), and  $\beta_1=0.20$  and  $\beta_2=0.05$ (dot-dashed curve).  For the central density of the star we have adopted the value $\rho_{c}=2.45\times 10^{15} {\rm g/cm}^3$, while $\bar{h}_0=0.1$ and $\bar{h}^\prime_0=0.5$. The stop-point in integration is $\rho=1.96\time10^{13}\, g/cm^3$.    }
	\label{BE-vec-v0}
\end{figure}

For a vanishing potential, with $V=0$, the behaviors of the mass and the density inside the BEC star is presented in Figs.~\ref{BE-mass-dens-v0}. The variation of  the non-zero component of the vector dark  energy $\bar{h}$ in terms of the $\eta$, inside and outside the BEC star is shown in Fig. \ref{BE-vec-v0}. The physical quantities behave in a proper manner.

\begin{figure}[htbp]
	\centering
	\includegraphics[width=8.0cm]{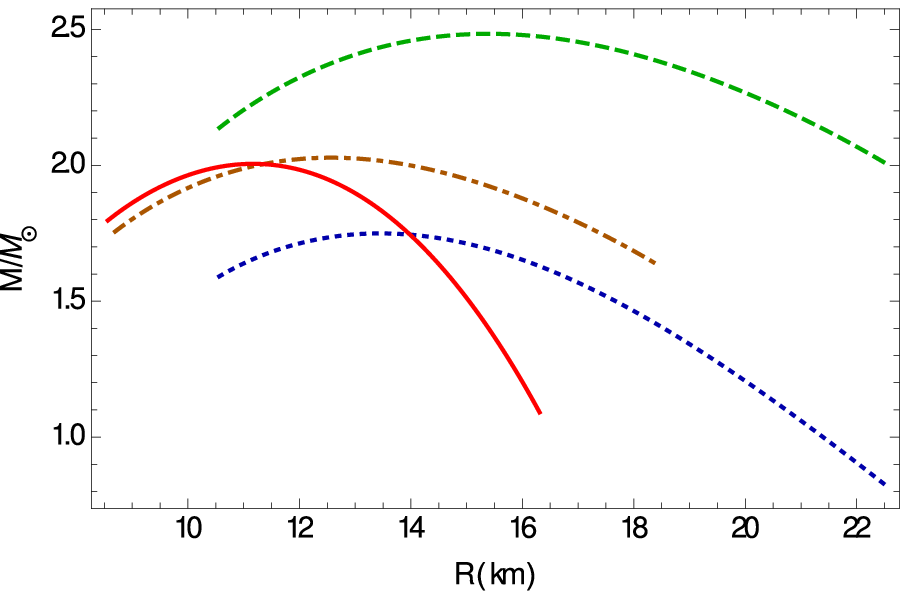}
	\caption{The mass-radius relation for BEC quark stars with $V=0$ for three different values of the	constants $\beta_1$ and $\beta_2$:  $\beta_1=-0.20$  and  $\beta_2=0.05$  (dashed curve),  $\beta_1=0.20$  and  $\beta_2=-0.01$  (dotted curve), and  $\beta_1=0.20$  and  $\beta_2=0.05$ (dot-dashed curve.   The solid curve represents the standard general relativistic  mass-radius relation for BEC stars.  }
	\label{BE-mr-v0}
\end{figure}

The mass-radius relation for this case is presented in Fig~\ref{BE-mr-v0}.

A few selected values of the maximum masses of BEC stars with $V=0$ are presented in Table~\ref{BE-v0-tab}. The maximum masses of the stars in the Einstein dark energy model exceed significantly the mass values of their general relativistic counterparts.
\begin{table}[h!]
	\begin{center}
		\begin{tabular}{|c|c|c|c|}
			\hline
			$\beta_1$ &~~~$-0.20$~~~&$~~~0.20~~~~$&$~~~0.20~~~~$ \\
			\hline
			$\beta_2$ &~~~$0.05$~~~&$~~~-0.01~~~~$&$~~~0.05~~~~$ \\
			\hline
			\quad$M_{max}/M_{\odot}$\quad& $~~~2.48~~~$& $~~~1.75~~~$& $~~~2.03~~~$\\
			\hline
			$~~~R\,({\rm km})~~~$& $~~~15.42~~~$& $~~~13.47~~~$& $~~~12.59~~~$\\
			\hline
			$~~~\rho_{c} \times 10^{-15}\,({\rm g/cm}^3)~~~$& $~~~1.74~~~$& $~~~2.74~~~$& $~~~1.74~~~$\\
			\hline
		\end{tabular}
		\caption{The maximum masses, and the corresponding radii  and central densities for BEC stars with $V=0$.}\label{BE-v0-tab}
	\end{center}
\end{table}

\subsubsection{The case $V=\lambda+a\Lambda_{\mu }\Lambda^{\mu }$}

\begin{figure*}[htbp]
	\centering
	\includegraphics[width=8.0cm]{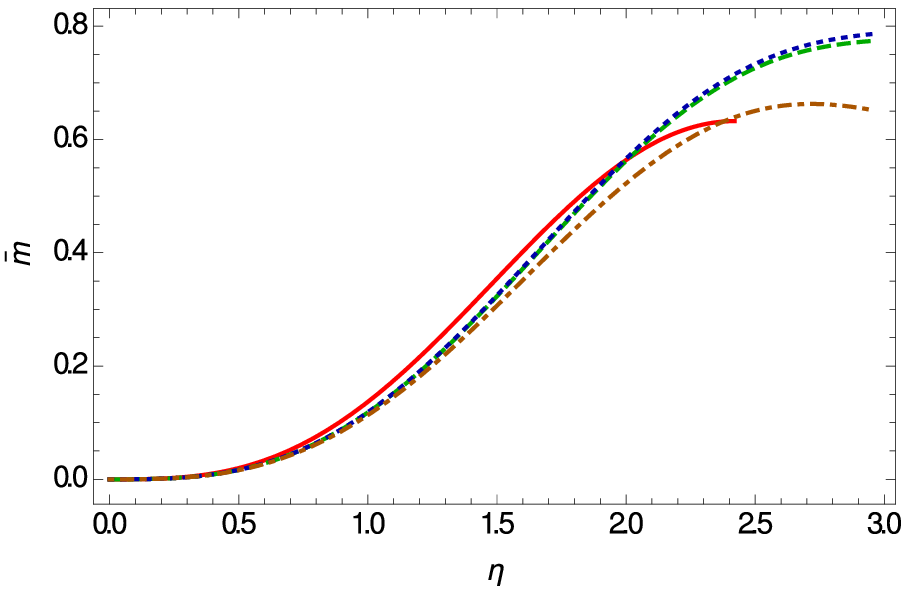}\hspace{.4cm}
	\includegraphics[width=8.0cm]{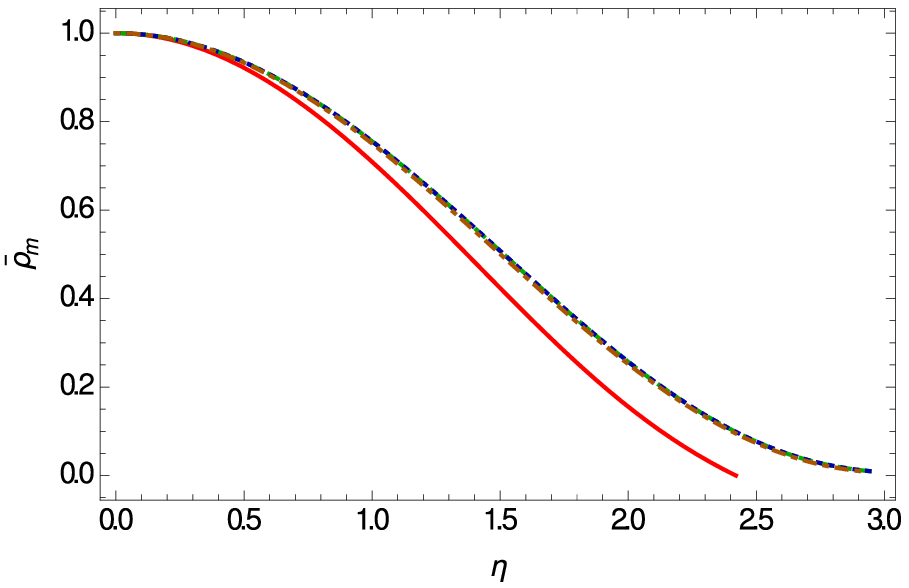}
	\caption{Variation of the interior mass and density profiles of BEC star with $V=\lambda+a\Lambda_{\mu }\Lambda^{\mu }$ as a function of the radial distance from the center of the  star $\eta$, for three different values of the
		constants $\bar{\lambda}$ and $\bar{a}$: $\bar{\lambda}=0$  and  $\bar{a}=-0.5$  (dashed curve),  $\bar{\lambda}=0.003$  and  $\bar{a}=0.0$  (dotted curve), and   $\bar{\lambda}=-0.03$  and  $\bar{a}=0.0$ (dot-dashed curve).  For the central density of the star we have adopted the value $\rho_{c}=2.45\times 10^{15} {\rm g/cm}^3$, while $\bar{h}_0=0.1$ and $\bar{h}^\prime_0=0.5$. The solid curve represents the standard general relativistic  mass and density profile for BE stars. }
	\label{BE-mass-dens-v}
\end{figure*}

\begin{figure}[htbp]
	\centering
	\includegraphics[width=8.0cm]{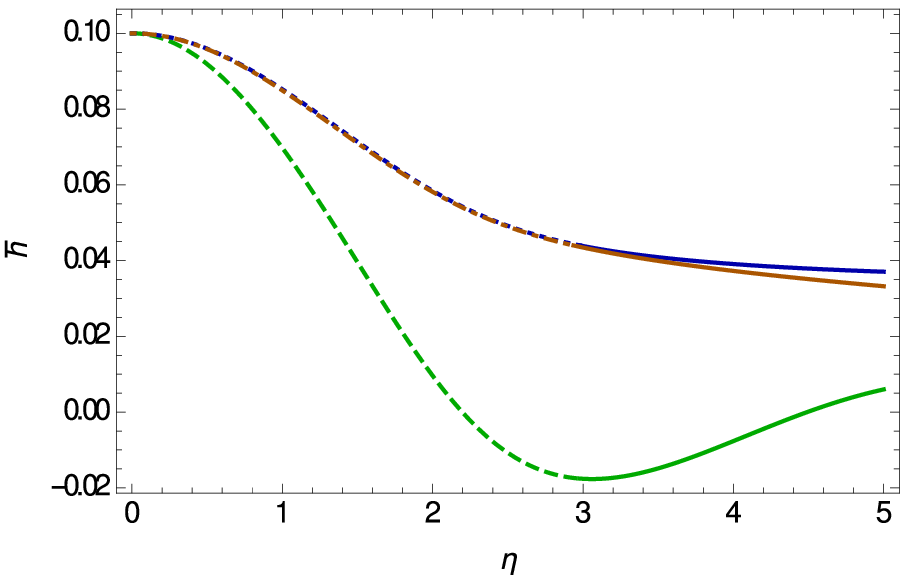}
	\caption{Variation of the temporal component of dark energy vector field inside and outside (solid lines) of the BEC star with $V=\lambda+a\Lambda_{\mu }\Lambda^{\mu }$ as a function of the radial distance from the center of the  star $\eta$, for three different values of the constants $\bar{\lambda}$ and $\bar{a}$: $\bar{\lambda}=0$  and  $\bar{a}=-0.5$  (dashed curve),  $\bar{\lambda}=0.003$  and  $\bar{a}=0.0$  (dotted curve), and   $\bar{\lambda}=-0.03$  and  $\bar{a}=0.0$ (dot-dashed curve).  For the central density of the star we have adopted the value $\rho_{c}=2.45\times 10^{15} {\rm g/cm}^3$, while $\bar{h}_0=0.1$ and $\bar{h}^\prime_0=0.5$.  }
	\label{BE-vec-v}
\end{figure}

Finally, we consider the case of the non-zero potential $V=\lambda+a\Lambda_{\mu }\Lambda^{\mu }$. For the coupling constants $\beta _1$ and $\beta _2$ we adopt the values $\beta_1=-0.20$ and  $\beta_2=0.05$, respectively, and we investigate the role of the different values of the parameters in the potential term.
The behaviors of the mass and density inside the BEC star is presented in Fig.~\ref{BE-mass-dens-v}. One can also see the effects of the values of the potential on the behavior of the vector dark energy $\bar{h}$ inside and outside of the BEC star in Fig \ref{BE-vec-v}. All the parameters of the star have a physical behavior.

\begin{figure}[htbp]
	\centering
	\includegraphics[width=8.0cm]{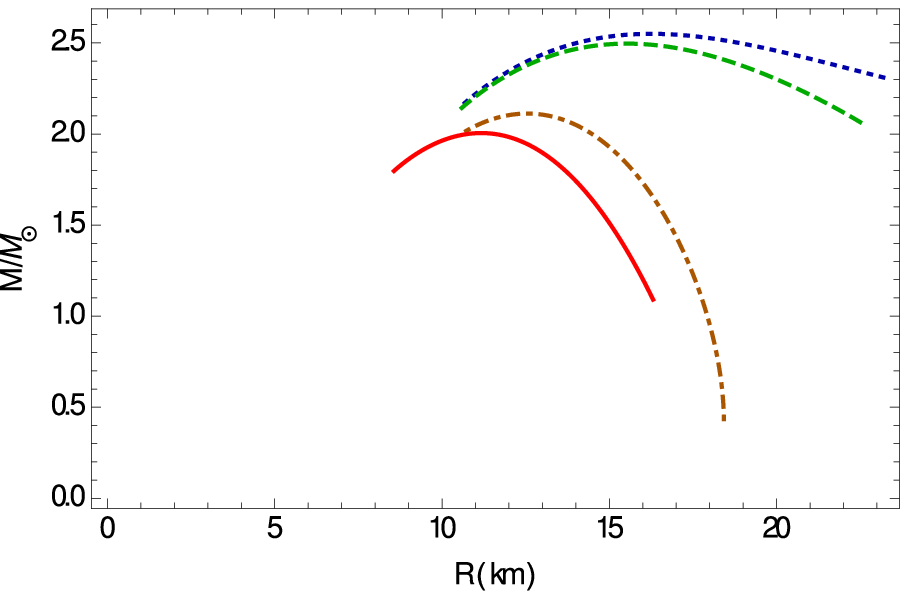}
	\caption{The mass-radius relation for BEC stars with $V=\lambda+a\Lambda_{\mu }\Lambda^{\mu }$ for three different values of the constants $\bar{\lambda}$ and $\bar{a}$: $\bar{\lambda}=0$  and  $\bar{a}=-0.5$  (dashed curve),  $\bar{\lambda}=0.003$  and  $\bar{a}=0.0$  (dotted curve), and   $\bar{\lambda}=-0.03$  and  $\bar{a}=0.0$ (dot-dashed curve).  For the central density of the star we have adopted the value $\rho_{c}=2.45\times 10^{15} {\rm g/cm}^3$, while $\bar{h}_0=0.1$ and $\bar{h}^\prime_0=0.5$.  The solid curve represents the standard general relativistic  mass-radius relation for BEC stars. }
	\label{BE-mr-v}
\end{figure}

The mass-radius relation for this case in shown in Fig.~\ref{BE-mr-v}. Several selected values of the maximum masses, and of the corresponding radii and central densities are presented in Table~\ref{BE-v-tab}.

\begin{table}[h!]
	\begin{center}
		\begin{tabular}{|c|c|c|c|}
			\hline
			$\bar{\lambda}$ &~~~$-0.03$~~~&$~~~0.0~~~~$&$~~~0.003~~~~$ \\
			\hline
			$\bar{a}$ &~~~$0.0$~~~&$~~~-0.5~~~~$&$~~~0.0~~~~$ \\
			\hline
			\quad$M_{max}/M_{\odot}$\quad& $~~~2.11~~~$& $~~~2.50~~~$& $~~~2.55~~~$\\
			\hline
			$~~~R\,({\rm km})~~~$& $~~~12.53~~~$& $~~~15.55~~~$& $~~~16.23~~~$\\
			\hline
			$~~~\rho_{c} \times 10^{-15}\,({\rm g/cm}^3)~~~$& $~~~3.45~~~$& $~~~1.70~~~$& $~~~1.49~~~$\\
			\hline
		\end{tabular}
		\caption{The maximum mass and corresponding radius  and central density for BEC stars with $V=\lambda+a\Lambda_{\mu }\Lambda^{\mu }$ for $\beta_1=-0.20$ and  $\beta_2=0.05$  .}\label{BE-v-tab}
	\end{center}
\end{table}

\subsection{Douchin-Haensel (SLy) type stars}

An equation of state of the neutron star matter, describing both the neutron star crust and the liquid core, was proposed in \cite{sly}. It was obtained by considering an effective nuclear interaction SLy of the Skyrme type, which is extremely useful for the calculation of the properties of very neutron rich matter. The structure of the crust, as well as its equation of state, are obtained by considering the zero temperature case, and by assuming the ground state composition. As for the crust-core transition, it is considered as a very weak first-order phase transition, with the relative density jump being of the order of one percent. The equation of state  of the liquid core is obtained by assuming that it consists of neutrons, protons, electrons, and muons only.
For this equation of state the minimum and maximum masses of static neutron stars are $0.094\,M_{\odot}$ and $2.05\,M_{\odot}$,
respectively.

In order to study the properties of neutron rich stars in the Einstein Dark Energy model we have adopted the data given in  Tables 3 and 5 of the paper \cite{sly} for the core and inner crust of the stars. The density of the core is between $1.30\times10^{14}\, {\rm g/cm}^3$ and $4.05\times 10^{15}\,{\rm g/cm}^3$. The density of the inner crust lies in the range
$3.49\times10^{11}\, {\rm g/cm^3}\leq\rho\leq1.28\times 10^{14}\,{\rm g/cm^3}$.  The equation of state for the outer crust is $p= K \rho^{4/3}$  \cite{sly}
where the constant $K$ will be determined by the continuity of the equation of state at the boundary between the inner and outer crusts.
In the following, two cases are considered. In the first case we assume that the potential $V$ vanishes, while in the second case we consider a quadratic type potential.  In both cases the results are compared to the ones obtained in standard general relativity. The stop point for integration in all cases is $\rho=3\times 10^{10}\;{\rm g/cm}^3)$. The initial condition for the vector field is $\bar{h}_0=0.1$ and $\bar{h}^\prime_0=0.5$.

\subsubsection{The case $V=0$}

\begin{figure}[htbp]
	\centering
	\includegraphics[width=8.0cm]{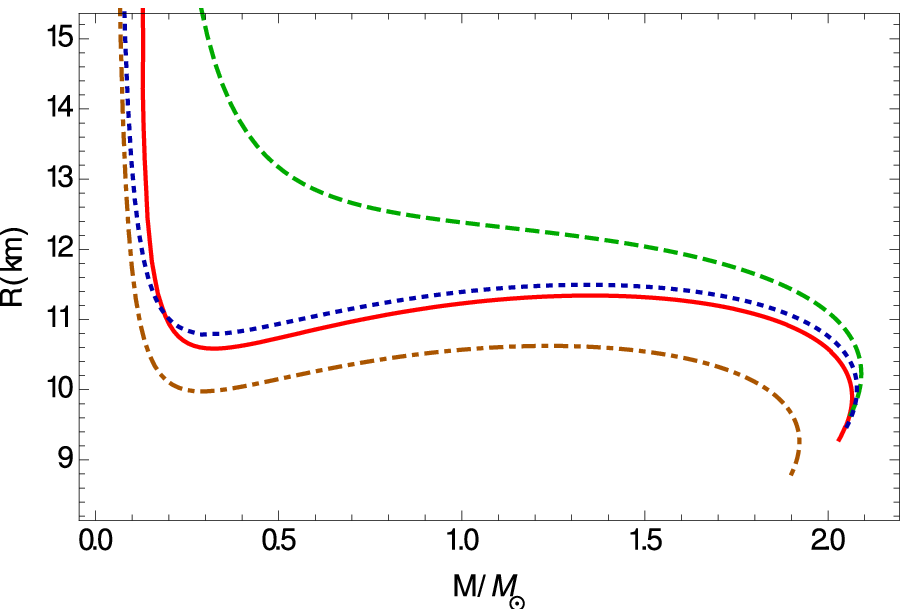}
	\caption{The radius-mass relation for SLy  stars for three different values of the	constants $\beta_1$ and $\beta_2$:  $\beta_1=-0.03$  and  $\beta_2=0.002$  (dashed curve),  $\beta_1=-0.03$  and  $\beta_2=-0.001$  (dotted curve), and  $\beta_1=0.12$  and  $\beta_2=-0.001$ (dot-dashed curve), respectively.   The solid curve represents the standard general relativistic  mass-radius relation.}
	\label{sly-mr-v0}
\end{figure}

The radius-mass relation for this case is shown in Fig.~\ref{sly-mr-v0}.

 The maximum masses, radii and central densities are presented, for some particular values of the model parameters $\beta _1$ and $\beta _2$,  in Table~\ref{sly-v0-tab}.

\begin{table}[h!]
	\begin{center}
		\begin{tabular}{|c|c|c|c|}
			\hline
			$\beta_1$ &~~~$~0.12~$~~~&$~~-0.03~~~$&$~~-0.03~~~$ \\
			\hline
			$\beta_2$ &~~~$-0.001$~~~&$~~~-0.001~~~~$&$~~~0.002~~~$ \\
			\hline
			\quad$M_{max}/M_{\odot}$\quad& $~~~1.92~~~$& $~~~2.08~~~$& $~~~2.09~~~$\\
			\hline
			$~~~R\,({\rm km})~~~$& $~~~9.25~~~$& $~~~10.01~~~$& $~~~10.23~~~$\\
			\hline
			$~~~\rho_{c} \times 10^{-15}\,({\rm g/cm}^3)~~~$& $~~~2.86~~~$& $~~~2.86~~~$& $~~~2.83~~~$\\
			\hline
		\end{tabular}
		\caption{The maximum masses, and the corresponding radii  and central densities for the SLy stars.}\label{sly-v0-tab}
	\end{center}
\end{table}

The presence of the vector field induces modifications in the basic stellar parameters of the neutron stars. In particular, positive values of the parameter $\beta _1$ lead to a decrease of the maximum mass of the star.

\subsubsection{The case $V=\lambda+a\Lambda_{\mu }\Lambda^{\mu }$}

\begin{figure}[htbp]
	\centering
	\includegraphics[width=8.0cm]{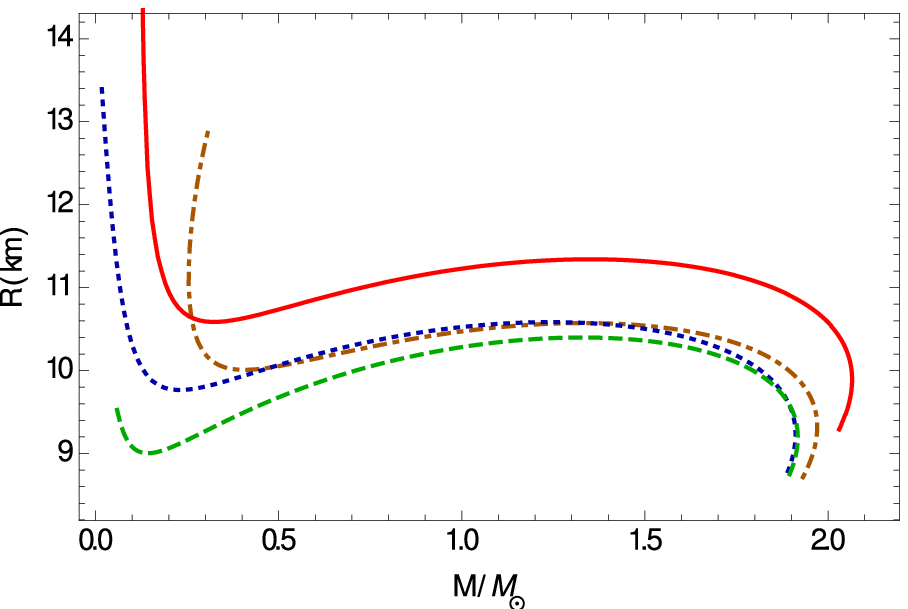}
	\caption{The radius-mass relation for SLy  stars for three different values of the	constants $\bar{\lambda}$ and $\bar{a}$:  $\bar{\lambda}=0$  and  $\bar{a}=0.3$  (dashed curve),  $\bar{\lambda}=-0.002$  and  $\bar{a}=0$  (dotted curve), and  $\bar{\lambda}=0.01$  and  $\bar{a}=0.3$ (dot-dashed curve), respectively.   The solid curve represents the standard general relativistic  mass-radius relation.  }
	\label{sly-mr-v}
\end{figure}

Now, we consider the case where the potential has the form $V=\lambda+a\Lambda_{\mu }\Lambda^{\mu }$. To investigate the numerical solution for this case we have set $\beta_1=0.12$  and  $\beta_2=-0.001$. The radius-mass relations for three different sets of the values of $\bar{\lambda}$ and $\bar{a}$ are represented in Fig.~ \ref{sly-mr-v}.

 The maximum masses, radii, and central densities for each cases are shown in Table~\ref{sly-v-tab}.

\begin{table}[h!]
	\begin{center}
		\begin{tabular}{|c|c|c|c|}
			\hline
			$\bar{\lambda}$ &~~~$0.0$~~~&$~~~-0.002~~~~$&$~~~0.01~~~~$ \\
			\hline
			$\bar{a}$ &~~~$0.3$~~~&$~~~0.0~~~~$&$~~~0.3~~~~$ \\
			\hline
			\quad$M_{max}/M_{\odot}$\quad& $~~~1.92~~~$& $~~~1.91~~~$& $~~~1.97~~~$\\
			\hline
			$~~~R\,({\rm km})~~~$& $~~~9.19~~~$& $~~~9.23~~~$& $~~~9.32~~~$\\
			\hline
			$~~~\rho_{c} \times 10^{-15}\,({\rm g/cm}^3)~~~$& $~~~2.90~~~$& $~~~2.90~~~$& $~~~2.69~~~$\\
			\hline
		\end{tabular}
		\caption{The maximum mass and corresponding radius  and central density for SLy stars for $\beta_1=0.12$ and  $\beta_2=-0.001$  .}\label{sly-v-tab}
	\end{center}
\end{table}

The variations of the potential parameters have a significant impact on the maximum mass of the star. For the considered values of the coupling parameters there is a slight decrease in the maximum masses of the stars, as compared to the $V=0$ case.

\section{Discussions and final remarks}\label{sect4}

In the present paper, we have investigated the basic physical properties of stellar type compact objects in the Einstein dark energy model,
 which brings together elements of the $f(R,T)$ gravity theory, and vector-tensor gravitational theories. Our main goal was to investigate if Einstein dark energy model can explain gravitational phenomena on both large cosmological scales, and astrophysical scales, through a single formalism. An important feature of the
theory is the presence of a vector type field, which plays an important role in the study of the interior of the dense compact
objects.

Vector type dark energy models, and their generalizations, in which dark energy is described by a vector field minimally coupled to gravity \cite{v1l,v1m}, a vector field non-minimally coupled to gravity \cite{v2,v2a,v2d}, or
by some extended vector field models \cite{v3,v4, S1,S2} has been intensively investigated in the literature. For example, the action for a
non-minimally massive vector field coupled to gravity can be introduced as \cite{v2}
\begin{align}
S =-&\int d^{4}x\sqrt{-g}\Bigg[\frac{R}{2}+\frac{1}{16\pi }F_{\mu \nu
}F^{\mu \nu }-\frac{1}{2}\mu _{\Lambda }^{2}A_{\mu }A^{\mu }  \notag
\label{v1} \\
&+\omega A_{\mu }A^{\mu }R+\eta A^{\mu }A^{\nu }R_{\mu \nu }+L_{m}\Bigg],
\end{align}%
where $A^{\mu }\left( x^{\nu }\right) $, $\mu ,\nu =0,1,2,3$ denotes the
four-potential of the dark energy, which  couples non-minimally
to gravity. Moreover, $\omega $ and $\eta $ are dimensionless coupling
parameters, while $\mu _{\Lambda }$ is the mass of the massive cosmological
vector field. The field tensor of the dark energy is given by $F_{\mu
\nu }=\nabla _{\mu }A_{\nu }-\nabla _{\nu }A_{\mu }$. Superconducting dark energy models that contain vector and scalar fields in
a gauge invariant way, were also investigated  \cite{S1, S2}.

In the present study we have concentrated on the local aspects of Einstein dark energy theory. From a physical point of view our main assumption is that the vector field $\Lambda _{\mu}$, which in Einstein's
theory can be interpreted from a cosmological point of view as vector type dark energy, plays also an important role at the stellar level.
 To obtain the field equations of the model, we have adopted a $%
f(R,T)$ type Lagrangian \cite{frt1}, which contains a linear combination of the Ricci
scalar, and of the trace of the energy-momentum tensor. Moreover, we construct
the self-interacting dark energy tensor field $C_{\mu \nu}$ in terms of the massive vector potential $\Lambda _{\mu}$. A coupling between the matter current and the vector potential can also be assumed, but in the present approach we have neglected this term.

It is important to point out that in the present gravitational action of the theory there is no direct (multiplicative) coupling between curvature and geometry, the action having an additive structure in the Ricci scalar and the trace of the matter energy-momentum tensor. However, the presence of $T$ and of the vector field in the action leads even in static spherically symmetry to a set of complicated interior field equations, which can be solved only by using numerical methods. We did begin  our study by deriving the basic field equations describing the structure of compact objects. From the field equations we have obtained the mass continuity equation,
the generalized TOV (hydrostatic equilibrium equation), and the equation of the vector field, given by a complicated second order nonlinear differential equation.  A physical/geometrical quantity that has an important effect in the determination of the properties of the
stellar objects is the self-interaction potential $V$ of the vector field. In our study we have assumed that $V$ either vanishes, $V=0$,
or it is quadratic in the vector field potential, thus having some similarities with the  Higgs potential of the scalar fields. Of course, different other choices of the potential are possible, and they will lead to dense stellar type object with different physical properties as compared to the properties of the stars analyzed in the present work.

Once the vector field self-interaction potential is fixed, to close the system of the structure equations of the star one must
specify the dense matter equation of state. We have adopted five equations of state  of the dense matter, and  we have constructed, through the numerical integration of the gravitational field equations, five classes of stellar models, corresponding to constant density stars, quark stars, CFL stars, Bose-Einstein condensate superfluid stars, and stars described by the Douchin-Haensel equation of state, respectively. For all these equations of state we have effectively obtained the Einstein dark energy model structure of the star, and compared it to its general relativistic counterpart. As a general conclusion of our study is that for all the five considered equations of state the Einstein dark energy model stars have a large variety of behaviors, determined by the variation of the model parameters. Much more massive stars than in
general relativity can also be obtained.  For example, if the maximum mass of a quark star in general relativity is of the order of $2M_{\odot}$, in the Einstein dark energy model the maximum mass of a quark star can have values of the order of $2.5M_{\odot}$ for $V=0$, and $2.8M_{\odot}$ for a quadratic self-interaction potential. For the CFL quark stars with $V\neq 0$ the maximum mass can reach values as high as $2.9M_{\odot}$, very close to the black hole stability limit of $3M_{\odot}$.  The masses of the stars are also strongly dependent on their central density. In the case of the Bose-Einstein condensate stars there is also an increase of the maximum mass from values of the order of $2M_{\odot}$ to masses in the range of $2.4-2.5M_{\odot}$.

A large number of high precision astronomical observations of the neutron star mass distribution have recently confirmed the existence of a large number of neutron stars
with masses of the order of $2M_{\odot}$, or higher \cite{Ma1,Ma2}. For example, the mass of the Black Widow Pulsar B1957+20, an eclipsing
binary millisecond pulsar, is estimated to be in the range $1.6-2.4M_{\odot}$ \cite{BW}. A range of $2-2.4M_{\odot}$
is very difficult to explain by using the standard neutron matter models together with general relativity, even if one admits the existence of exotic
particles inside the stars, including quarks or kaons. But these values of the stellar masses can be explained rather easily once we assume they are
Einstein dark energy model stars. As we have seen, an Einstein dark energy model star has an internal structure that is more complex than that of the general relativistic stars.

One important question is if and how different types of stellar models can be distinguished observationally. One such possibility is related to the study of accretion disks that form around massive stellar type objects \cite{accr1,accr2, accr3,accr4}. As a result of the differences in the exterior geometry (metric), the thermodynamic and electromagnetic properties of the disks (temperature distribution, energy flux, equilibrium radiation spectrum, and efficiency of energy conversion, respectively) are different for different classes of dense stellar type objects. Therefore,  the emissivity properties of the accretion disks, and of the compact objects themselves, could be the key signature that would allow to differentiate Einstein dark energy model stars from compact general relativistic objects.
However, even if a number of distinctive astrophysical signatures that could differentiate between different classes of stars may exist at a theoretical level, their observational detection may prove to be an extremely difficult task. The possible observational/astrophysical
features of the Einstein dark energy stars will be discussed in a future publication.

\section*{Acknowledgments}

The work of T. H. is supported by a grant of the Romanian Ministry of Education and Research, CNCS-UEFISCDI, project number PN-III-P4-ID-PCE-2020-2255 (PNCDI III).

\end{document}